\documentclass{aa}

\usepackage[T1]{fontenc}
\usepackage{pslatex}
\usepackage{empheq}
\usepackage{amsmath}
\usepackage{mathabx}
\usepackage{enumitem}
\usepackage{amsfonts}
\usepackage{xcolor}
\usepackage{url}
\usepackage{bm}
\usepackage{graphicx}
\usepackage{amssymb}
\usepackage{verbatim}
\usepackage{soul}
\usepackage{enumitem}
\usepackage{booktabs}
\usepackage{multirow}
\usepackage{array}
\usepackage[utf8]{inputenc}
\inputencoding{latin1}
\inputencoding{utf8}
\usepackage{appendix}

\def\lsim{ \lower .75ex\hbox{$\sim$} \llap{\raise .27ex \hbox{$<$}} }
\def\gsim{ \lower .75ex \hbox{$\sim$} \llap{\raise .27ex \hbox{$>$}} }

\newcommand{\bi}{\begin{itemize}}
\newcommand{\ei}{\end{itemize}}

\usepackage[colorlinks=true,linkcolor=blue,citecolor=blue]{hyperref}
\begin{document}

\title{Temporal variability of polarization in blazars}

\titlerunning{Polarization in blazars}

\author{
F. Bolis\inst{1,2}
\and E. Sobacchi\inst{3,2}
\and F. Tavecchio\inst{2}
}
\authorrunning{Bolis et al.}

\institute{
DiSAT, Università dell’Insubria, Via Valleggio 11, I-22100 Como, Italy\\
\email{filippo.bolis@inaf.it}
\and 
INAF -- Osservatorio Astronomico di Brera, Via E. Bianchi 46, I-23807 Merate, Italy
\and
Gran Sasso Science Institute, Viale F.~Crispi 7, I-67100 L’Aquila, Italy
}
\date{}

\voffset-0.4in



\abstract{
We investigate the temporal variability of polarization of synchrotron radiation from blazar jets. Multiwavelength observations revealed high-amplitude rotations of the electric vector position angle (EVPA), both in the optical and in the X-rays. More often, the polarization degree and the EVPA show a seemingly erratic variability. To interpret these observations, we present a geometric and deterministic model in which off-axis, compact emitting features (i.e.,~blobs) propagate along the jet with the local velocity of the flow. The dynamics of the blobs is determined by the jet electromagnetic fields, which are calculated self-consistently using an analytical model of magnetically dominated outflows. The jet is axisymmetric, and its electromagnetic fields do not have a turbulent component. We show that the observed polarization is sensitive to the initial spatial configurations of the blobs. For the same jet structure, we observe several remarkably complex polarization patterns, including large EVPA rotations of $180^{\circ}$ or more in both directions and more erratic fluctuations. Simultaneous high-amplitude variations of the polarization degree and the EVPA can coincide with peaks of the observed luminosity. However, seemingly uncorrelated variations are also possible. We discuss the feasibility of constraining the particle acceleration mechanism from multifrequency polarimetric observations.
}


\maketitle
\boldsymbol{}

\section{Introduction}

Relativistic jets from supermassive black holes (SMBHs) in active galactic nuclei (AGNs) are highly collimated outflows of magnetized plasma. Charged particles produce non-thermal radiation that spans the entire electromagnetic spectrum, from the radio band to very high-energy gamma-rays \citep{Blandford19}. Blazars, a subclass of AGNs where the jet is closely aligned to our line of sight, are optimal probes of the jet physics, as the non-thermal radiation is strongly beamed due to the advantageous orientation \citep{romero17, Boettcher19}.

The broadband spectral energy distribution (SED) of blazars shows a double-humped shape that peaks in the IR-soft-X-ray band and in the MeV-TeV band \citep[e.g.,][]{fossati98}. The low-energy part of the SED is attributed to synchrotron radiation emitted by non-thermal electrons. Depending on the frequency at which the synchrotron component peaks, $\nu_{\rm peak}$, blazars can be classified as high-synchrotron-peaked (HSP), intermediate-synchrotron-peaked (ISP), and low-synchrotron-peaked (LSP) \citep[e.g.,][]{Ajello2022}. HSP blazars have $\nu_{\rm peak} > 10^{15} \; \rm{Hz}$, ISP blazars have $10^{14}\; \rm{Hz}<\nu_{\rm peak} < 10^{15}\; \rm{Hz}$, and LSP blazars have $\nu_{\rm peak} < 10^{14}\; \rm{Hz}$. Most HSP and ISP blazars are classified as BL Lac objects, whereas LSP blazars include both Flat Spectrum Radio Quasars (FSRQs) and BL Lacs.

Since synchrotron radiation (which is responsible for the low-energy part of the SED) is intrinsically linearly polarized, polarimetry has attracted significant attention as a key diagnostic of blazar jets. Observations of blazars in the quiescent state show that the optical polarization degree, $\Pi_{\rm O}$, is between ${\rm a \; few\;}\%$ and $30\%$. In HSP blazars, where both optical and X-ray emission is due to synchrotron radiation, the X-ray polarization degree is significantly higher than the optical one: $\Pi_{\rm X} / \Pi_{\rm O} \,\gsim \, 2$ \citep[e.g.,][]{liodakis22, digesu22, Kouch2024}.

Episodes of optical polarization variability in blazars have been reported during the \textit{Robopol} monitoring campaigns \citep{pavlidou14}. These observations revealed that both the polarization degree and the electric vector position angle (EVPA) can show significant variations, especially during active states \citep[e.g.,][]{Marscher08, marscher10, blinov16, Blinov18}. A notable feature is that the EVPA can undergo substantial rotations, or ``swings'' \citep{Marscher08, marscher10, Abdo2010b, Larionov2013, blinov15, blinov16, Blinov18, Kiehlmann2016}. EVPA rotations with amplitudes $\gtrsim 90^{\circ}$ are relatively rare events, although even larger swings have been reported \citep{Marscher08, marscher10, chandra2015}. The largest EVPA rotation detected to date is $\sim 720^{\circ}$ in PKS 1510-089 \citep{marscher10}. The timescales of EVPA rotations range from weeks to a few hours \citep{marscher10, blinov15}. Interestingly, the EVPA can rotate in both directions within the same source \citep{chandra2015}.

EVPA swings are often accompanied by $\gamma$-ray flares and by a temporary decrease of the optical polarization degree \citep{blinov15, blinov16, Blinov18, Kiehlmann2016}. The fact that wide rotations are connected with multiwavelength flares \citep{Abdo2010, marscher10, Aleksic2014} was interpreted as evidence that EVPA swings are produced by a deterministic mechanism \citep{Blinov18}. However, the overall temporal polarization pattern often appears erratic, which could be the signature of a stochastic process \citep{marscher14, Kiehlmann2017}.

The recent launch of the Imaging X-ray Polarimetry Explorer satellite \citep[{\it IXPE},][]{weisskopf22} extended polarimetric observations to the X-ray domain, making it possible to perform multiwavelength polarimetric campaigns \citep{liodakis22, digesu22, digesu23, ehlert2023, middei23b, middei23a, peirson23, Chen2024, Errando2024, Kim2024, Kouch2024, Marshall2024, Abe2025, Agudo2025, Liodakis2025, Lisalda2025, Pacciani2025}. Several \textit{IXPE} observations target HSP blazars, where X-rays are produced through synchrotron radiation. In principle, polarimetric observations of HSP blazars can reveal the magnetic field structure within the acceleration site \citep{tavecchio21}.

\textit{IXPE} detected EVPA swings in the X-ray band, which are not associated with simultaneous optical swings \citep{digesu23, Kim2024}. These observations are interpreted as evidence that electrons are accelerated by a shock moving on a helical trajectory \citep{digesu23}. EVPA swings in the X-ray band arise because the emitting electrons illuminate a small region of the jet in the shock downstream. Instead, the optical EVPA does not rotate because the emitting electrons are distributed more uniformly as a result of their longer cooling time. However, the energy-stratified shock model can hardly explain the optical EVPA swings without a simultaneous swing in the X-ray band \citep{middei23a}. We emphasize that energy stratification of the emitting electrons is not a unique signature of the shock acceleration scenario, but is a generic by-product of any spatially localized acceleration process \citep[][]{Bolis+2024, Zhang+2024}.

The discovery of variability in polarization sparked a number of theoretical studies aimed at interpreting this behavior \citep[for a review, see][]{H.Zhang2019}. EVPA swings have been attributed to geometric effects \citep[e.g.,][]{Marscher08, marscher10, Nalewajko2010, Larionov2013, Lyutikov2017, Peirson2018} or to physics effects including changes in the magnetic field structure due to shocks \citep{Koenigl85, Zhang2014, Zhang2015,Zhang2016b} and current-driven kink instabilities \citep{Nalewajko2017, Zhang2016}.
Random walk processes have been invoked to interpret the erratic variability of polarization \citep{Kiehlmann2017}. Scenarios in which both deterministic and stochastic mechanisms coexist in the same source (e.g.,~shock waves in turbulent magnetic fields) were also considered \citep{marscher14, angelakis16, tavecchio18}.
Recent fully-kinetic particle-in-cell (PIC) simulations show that EVPA swings may be the results of magnetic reconnection events \citep{Zhang2018, HoskingSironi2020}. In the reconnection scenario, the EVPA swings would be associated with high-energy outbursts.

\begin{figure}
\centering
\includegraphics[width=0.45\textwidth]{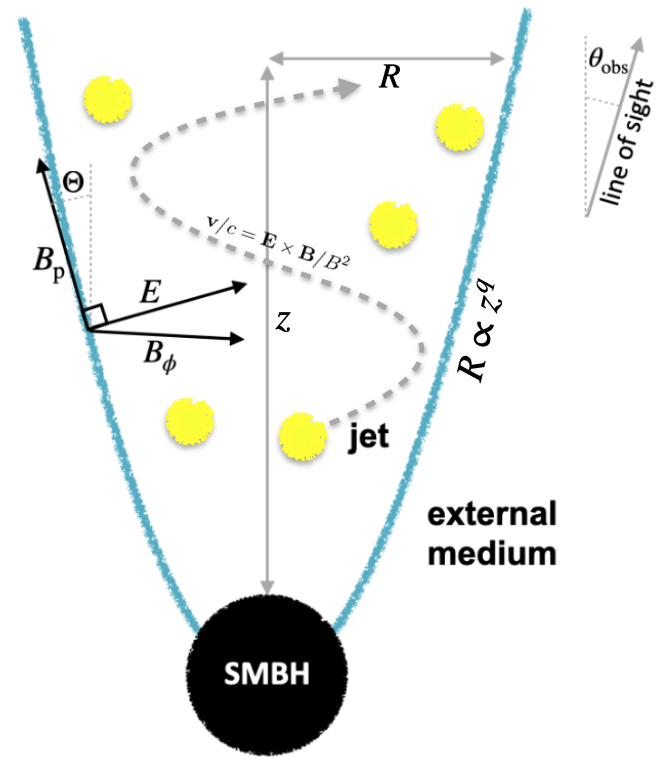}  
\caption{Cartoon of our jet model. The equilibrium boundary radius of the jet, $R$, is proportional to $z^q$, where $z$ is the distance from the supermassive black hole. The value of the parameter $q<1$ depends on the pressure profile of the external medium that collimates the jet. We consider a cylindrical jet ($q=0$), a ``sausage-like'' jet ($q=0.3$), and a nearly parabolic jet ($q=0.4$). In the ``sausage-like'' jet, the boundary oscillates about the equilibrium value. The emission is produced by an ensemble of compact off-axis emission features (yellow blobs) that propagate along the jet with the local velocity of the flow. The line of sight makes an angle $\theta_{\rm obs}$ with respect to the direction of the jet axis.}
\label{fig:scenario}
\end{figure}

In this work, we propose a geometric and deterministic model to interpret the variability in polarization. Following the approach of \cite{Marscher08, marscher10} and \cite{Larionov2013}, we consider compact off-axis emitting features (i.e.,~blobs) propagating along the jet with the local velocity of the flow (see Fig.~\ref{fig:scenario}). The key aspect of our model is that, unlike previous approaches, the blob's trajectory is computed self-consistently. Its dynamics is dictated by the structure of jet electromagnetic fields, whose profile is derived from a model of magnetically driven outflows \citep{Lyubarsky2009}. We do not assume a specific acceleration mechanism for the emitting particles. The variability in polarization is determined by purely kinematics effects (i.e.,~relativistic beaming and light-travel time delays). We find that even a geometric and deterministic model can produce remarkably complex polarization patterns, including large EVPA rotations and seemingly erratic variability.

The paper is organized as follows. In Sect.~\ref{sec:jetmodel}, we describe the jet model adopted in this work. In Sect.~\ref{sec:res}, we present our results. In Sect.~\ref{sec:disc}, we discuss their implications and summarize our conclusions. 

\section{Physical model}
\label{sec:jetmodel}

According to the leading theoretical model, relativistic jets are propelled by the extraction of magnetic energy from rotating SMBHs \citep[e.g.,][]{BlandfordZnajek1977, Vlahakis&Konigl2004, Tchekhovskoy2011}. Jet power is still dominated by Poynting flux in the region where optical and X-ray emissions are produced \citep{Lyubarsky2010}. The structure of magnetically dominated jets has been extensively studied using both analytical methods \citep{Beskin+1998, Beskin+2004, Vlahakis2004, Lyubarsky2009, Lyubarsky2010} and numerical simulations \citep{Komissarov2007, Komissarov2009, Tchekhovskoy09, Tchekhovskoy2011}.

\subsection{Jet structure}

Our jet model is based on the work of \cite{Lyubarsky2009}, who derives asymptotic equations (valid far from the light cylinder) to describe axisymmetric, stationary, Poynting-dominated MHD outflows. The external pressure that confines the outflow decreases as a power-law, $\mathcal{P}_{\rm ext}\; \propto\; z^{-\kappa}$, where $z$ denotes the distance from the SMBH. When $\kappa \leq 2$, the jet has a parabolic shape, $R\; \propto\; z^{q}$, where $R$ is the transverse radius of the jet and the parameter $q < 1$ depends on the power-law index $\kappa$ of the external pressure profile.
For $0 < \kappa < 2$, one has $q = \kappa/4$, while for a wind-like medium with $\kappa = 2$, one has $1/2 < q < 1$. Throughout this work, we focus on solutions with $\kappa < 2$, and therefore $q < 1/2$. We adopt cylindrical coordinates $(R, \phi, z)$, with the $z$ axis aligned with the jet direction. Quantities without a prime are defined in the observer frame, while primed quantities are defined in the jet frame. For simplicity, we assume that the SMBH is located at redshift $z=0$. 

The electromagnetic fields in the observer frame can be written as 
\begin{align}
\label{eq:Efield}
\mathbf{E} & = E_{R} \hat{\mathbf{R}} + E_{z} \hat{\mathbf{z}}\\
\label{eq:Bfield}
\mathbf{B} & = B_{R} \hat{\mathbf{R}} + B_{\phi} \hat{\bm{\phi}} +B_{z} \hat{\mathbf{z}} \,.
\end{align}
The components of the electromagnetic fields are given by
\begin{align}
\label{eq:Ecomp}
E_{R} & = \Omega R B_{\rm p} \cos \Theta\,, & E_{z} & = -\Omega R B_{\rm p} \sin \Theta \\
\label{eq:Bcomp}
B_{R} & = B_{\rm p}\sin \Theta\,, & B_{z} & = B_{\rm p} \cos \Theta \,,
\end{align}
where $\Theta$ is the local opening angle of the jet. We adopt units in which the speed of light is set to $c = 1$. The angular velocity $\Omega$ and the poloidal magnetic field $B_{\rm p}$ are assumed to be independent of $R$. Numerical simulations support this approximation (see, e.g., Fig.~4 of \citealt{Komissarov2007}).

The shape of the magnetic flux surfaces (where the flux function $\psi$ is constant) is given by Eq.~(74) of \citet{Lyubarsky2009}, which is
\begin{equation}
\label{eq:profile}
    \Omega R  = 3^{1 / 4} \left(\frac{\psi}{ \psi_0}\right)^{1/ 2} Y\left( \Omega z \right)\,.
\end{equation}
The dimensionless function $Y(\Omega z)$ is defined by Eqs.~(81)-(82) of \citet{Lyubarsky2009} as
\begin{equation}
\label{eq:Y}
    Y= K \,\left(\Omega z\right)^q \Bigg[ \frac{1}{C_1} \cos^2 S + C_1 \Big( C_2\cos S + \frac{\pi}{2-4q} \sin S \Big)^2\Bigg]^{1/2}\, , 
\end{equation}
where
\begin{equation}
  \label{eq:S}  
  S = \frac{\sqrt{\beta}}{1-2q} \, \left(\Omega z\right)^{1-2q}  - \frac{1-q}{1-2q}\,\frac{\pi}{2}\;, \quad   {\rm and} \quad K= \sqrt{\frac{2-4q}{\pi\, \beta^{1/2}}}\,. 
\end{equation}
In the above equations, $\beta\sim 1$ denotes the ratio of the external pressure to the magnetic pressure at the light cylinder, while $C_1$ and $C_2$ are free parameters.

The local opening angle of the jet, $\Theta$, can be expressed as \citep{Lyubarsky2009, Bolis+2024}
\begin{equation}
\label{eq:thetajet}
\Theta \equiv \left|\frac{ \partial \psi/\partial z }{\partial \psi/\partial R } \right| = \frac{R}{Y} \frac{{\rm d}Y}{{\rm d}z} = 3^{1 / 4} \left(\frac{\psi}{ \psi_0}\right)^{1/ 2}\frac{{\rm d}Y}{{\rm d}\left(\Omega z\right)} \,,
\end{equation}
and the toroidal magnetic field is given by
\begin{equation}
\label{eq:BminusE}
 \frac{B^2_\phi - E^2}{B^2_{\rm p}} = - \left(\frac{\psi}{ \psi_0}\right)^2 Y^3 \, \frac{{\rm d}^2Y}{{\rm d}\left(\Omega z\right)^2} \,.
\end{equation}
The poloidal magnetic field is given by Eq.~(5) of \citet{Lyubarsky2009}, which is ${\bm B}_{\rm p}= \nabla \psi \times \hat{\bm{\phi}}/R$. Its magnitude is
\begin{equation}
B_{\rm p} = \frac{2\,\psi_0}{\sqrt{3}\,Y^2}\,
\sqrt{1+\sqrt{3} \left(\frac{\psi}{ \psi_0}\right)\left[ \frac{{\rm d} Y}{{\rm d} \left(\Omega z\right)}\right]^2}\;.
\end{equation}
Since the outflow is magnetically dominated, we assume that the bulk velocity of the fluid, $\mathbf{v}$, coincides with the drift velocity: $\mathbf{v}=\mathbf{E}\times\mathbf{B}/B^2$. The corresponding bulk Lorentz factor is $\Gamma=(1-v^2)^{-1/2}=(1-E^2/B^2)^{-1/2}$.

\begin{figure*}
    \centering
    \includegraphics[width=0.32\textwidth]{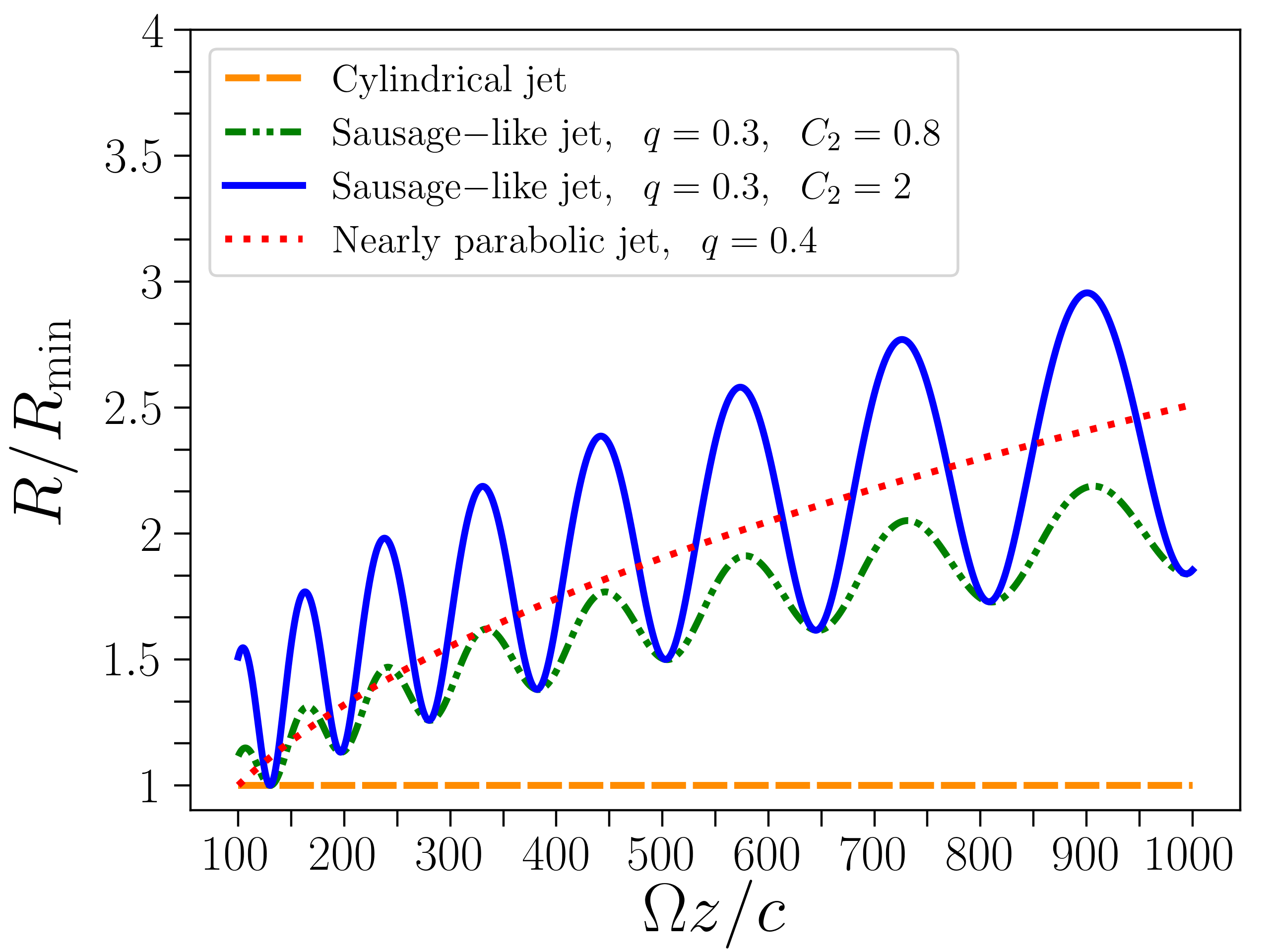}
    \hspace{+0.2em}
    \includegraphics[width=0.32\textwidth]{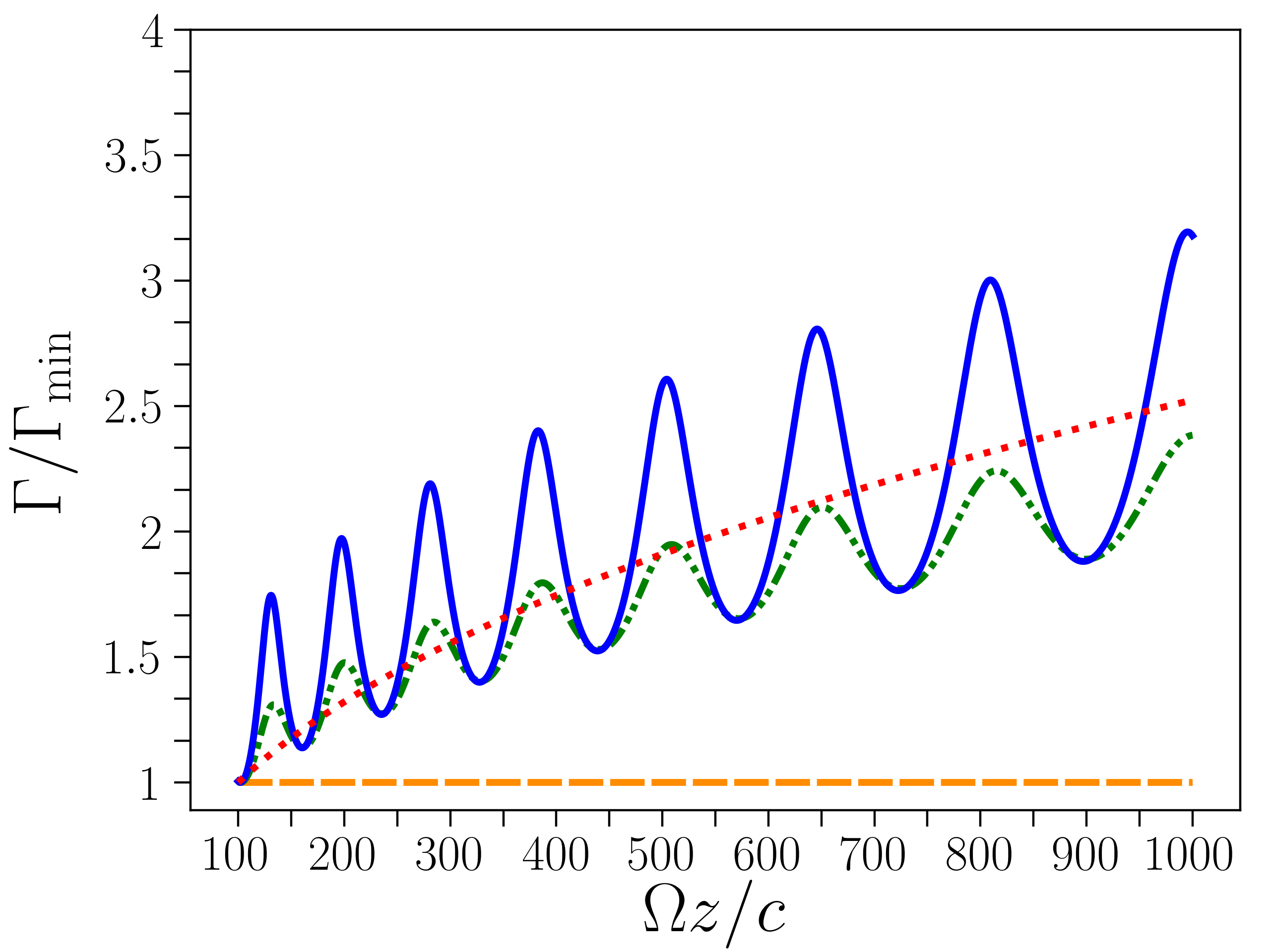}
    \hspace{+0.2em}
    \includegraphics[width=0.32\textwidth]{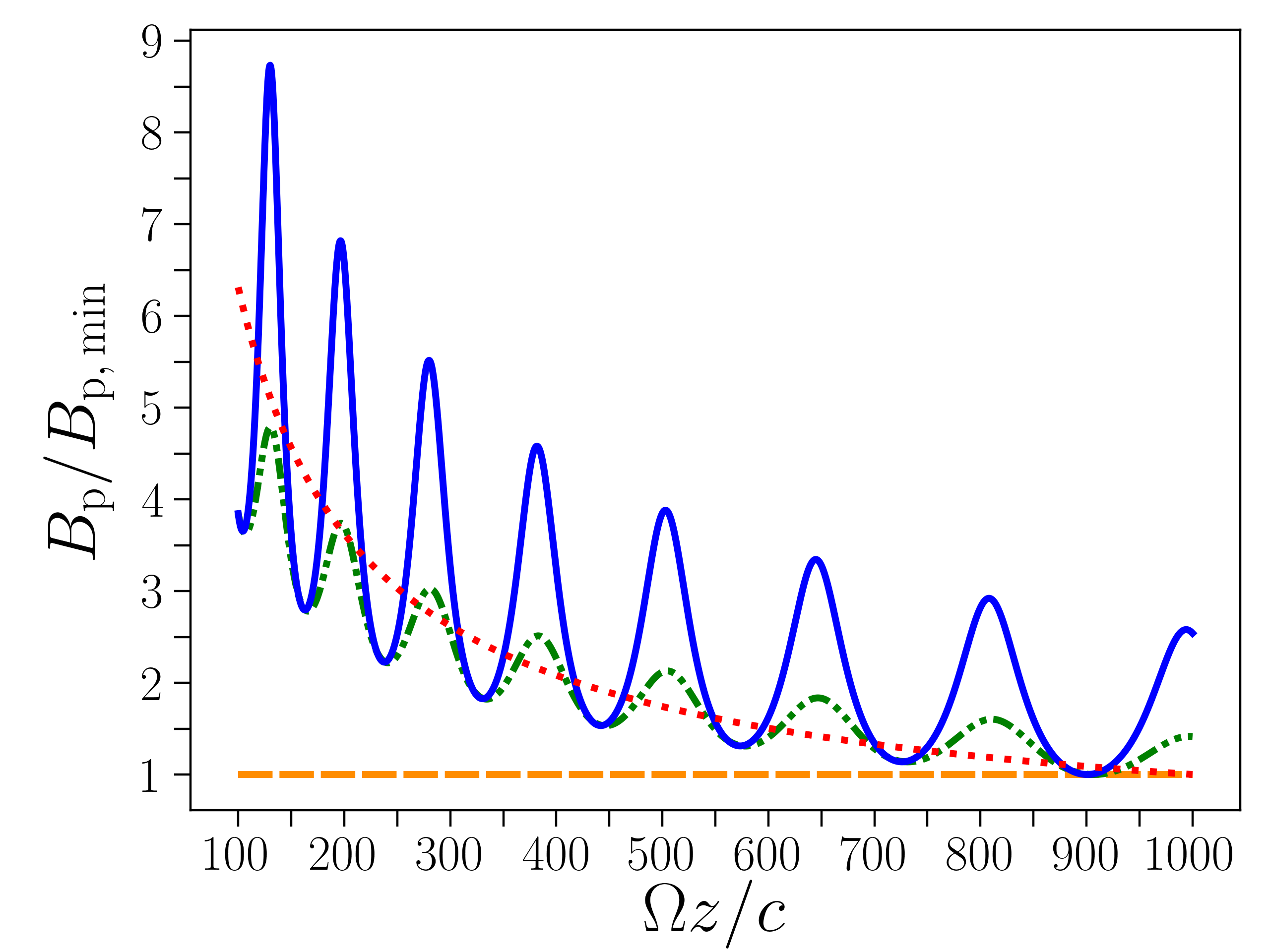}
    \caption{
    Jet transverse radius normalized to its minimum value, $R / R_{ \mathrm{min}}$ (left panel), bulk Lorentz factor in the proper frame normalized to its minimum value, $\Gamma / \Gamma_{\mathrm{min}}$ (middle panel), and poloidal magnetic field magnitude normalized to its minimum value, $B_{\rm p}/ B_{\rm{p,\, min}}$ (right panel), as functions of the distance from the black hole expressed in units of $c/\Omega$, for different jet shapes.}
    \label{fig:RandGamma}
\end{figure*}

\subsection{Time-dependent polarimetry of moving blobs}
\label{sec:timepol}

In the fluid proper frame, the distribution of emitting electrons is isotropic in momentum and follows a power-law in energy:
\begin{equation}
    \frac{dN_e}{d\gamma_e} = K_{\rm e}\,\gamma^{-p}_e \, ,
\end{equation}
where $K_{\rm e}\big(t,R,\phi, z \big)$ is the proper number density, $\gamma_{\rm e}$ is the Lorentz factor of the electrons, and $p$ is the power-law index of the electrons.

We determine the time-dependent polarization of synchrotron radiation as follows. We add emission from an ensemble of bright off-axis compact features (blobs) that propagate with the local bulk velocity of the fluid. We assume that all the blobs are identical.
The calculation of the Stokes parameters ($I$, $Q$, $U$) of a single blob is described in detail in Appendix \ref{sec:StokesT}. Since we are dealing with ultra-relativistic particles, we neglect circular polarization and set $V=0$ (where $V$ is the fourth Stokes parameter). The normalized Stokes parameters of the ensemble of blobs are defined as
\begin{align}
\label{eq:Iblobs}
I_{\rm blobs} & = \frac{1}{N_{\rm blobs}} \sum_{i=1}^{N_{\rm blobs}}I_i\\
Q_{\rm blobs} & = \frac{1}{N_{\rm blobs}} \sum_{i=1}^{N_{\rm blobs}}Q_i\\
\label{eq:Ublobs}
U_{\rm blobs} & = \frac{1}{N_{\rm blobs}} \sum_{i=1}^{N_{\rm blobs}}U_i \;,
\end{align}
where the index $i$ refers to the $i$-th blob, and $N_{\rm blobs}$ is the total number of blobs. We consider a scenario with a single blob ($N_{\rm blobs}=1$) and a scenario with multiple blobs (in this case, we assume $N_{\rm blobs}=10$). Multiple blobs could be produced if particles are accelerated at different reconnection sites. We have verified that our results remain qualitatively similar if $N_{\rm blobs}$ increases or decreases by a factor of a few with respect to our fiducial value ($N_{\rm blobs}=10$). The normalization of the Stokes parameters is irrelevant for the calculation of the polarization degree and the EVPA, which depend on the ratios $Q_{\rm blobs}/I_{\rm blobs}$ and $U_{\rm blobs}/I_{\rm blobs}$. With the normalization of Eqs.~\eqref{eq:Iblobs}-\eqref{eq:Ublobs}, the total radiation intensity is independent of $N_{\rm blobs}$. Changing $N_{\rm blobs}$ is equivalent to modifying the spatial distribution of the emitting particles, while keeping their total number the same.

We consider the possibility that part of the synchrotron radiation is produced by a nearly axisymmetric jet. We model the jet as an additional ensemble of $N_{\rm jet}$ blobs that propagate with the local bulk velocity of the fluid. The normalized Stokes parameters of the jet are defined as
\begin{align}
I_{\rm jet} & = \frac{1}{N_{\rm jet}} \sum_{i=1}^{N_{\rm jet}}I_i\\
Q_{\rm jet} & = \frac{1}{N_{\rm jet}} \sum_{i=1}^{N_{\rm jet}}Q_i\\
\label{eq:Ujet}
U_{\rm jet} & = \frac{1}{N_{\rm jet}} \sum_{i=1}^{N_{\rm jet}}U_i \;.
\end{align}
We assume that $N_{\rm jet}$ is very large (i.e.,~$N_{\rm jet}\gg N_{\rm blobs}$). Then, the normalized Stokes parameters of the jet are independent of the exact value of $N_{\rm jet}$ because the distribution of the blobs is nearly axisymmetric (we find that $N_{\rm jet}=10^5$ is sufficient for our scope).

The final values of the Stokes parameters are obtained by adding the blobs and the jet:
\begin{align}
I & = I_{\rm blobs} + \eta_{\rm jet}\, I_{\rm jet} \\
Q & = Q_{\rm blobs} + \eta_{\rm jet}\, Q_{\rm jet} \\
U & = U_{\rm blobs} + \eta_{\rm jet}\, U_{\rm jet}\;,
\end{align}
where $\eta_{\rm jet}$ is a weighting factor. We consider a scenario where the axisymmetric jet is absent ($\eta_{\rm jet}=0$), and a scenario where the jet and the blobs have similar luminosities\footnote{One could define the Stokes parameters without the normalization used in Eqs.~\eqref{eq:Iblobs}-\eqref{eq:Ujet}, namely $\{ I_{\rm blobs}, Q_{\rm blobs}, U_{\rm blobs}\} =\sum_{i=1}^{N_{\rm blobs}}\{ I_i, Q_i, U_i\}$ and $\{ I_{\rm jet}, Q_{\rm jet}, U_{\rm jet}\} =\sum_{i=1}^{N_{\rm jet}}\{ I_i, Q_i, U_i\}$. With this definition, the scenario where the jet and the blobs have similar luminosities would correspond to $\eta_{\rm jet}=N_{\rm blobs}/N_{\rm jet}$.} ($\eta_{\rm jet}=1$). The polarization degree, $\Pi$, and the EVPA, $\Psi$, are given by
\begin{equation}
\label{PI}
\Pi = \frac{\sqrt{Q^{2} + U^{2}}}{I}\;,\qquad
\tan 2 \Psi = \frac{U}{Q} \,.
\end{equation}

\subsection{Parameters of the model}

We assume that the viewing angle is $\theta_{\rm obs} = 0.1{\rm \, rad}$ and that the angular velocity is $\Omega = 10^{-5} {\rm \, s^{-1}}$. Our choice gives $\Gamma\sim 10$ at the distance from the SMBH where the non-thermal emission is produced ($z\sim 10^{17}-10^{18}{\rm\; cm}$). We assume that the blobs are located near the boundary radius of the jet (i.e., on the surface that separates the jet from the external medium). Then we set $\psi = \psi_0$ in Eqs.~\eqref{eq:profile}-\eqref{eq:BminusE}. Our choice is appropriate if the particles are accelerated by shear flow near the boundary radius \citep[e.g.,][]{Sironi2021}.

In the scenario with a single blob ($N_{\rm blobs}=1$), the coordinates of the blob at the initial time $t=0$ (namely, $\phi_0, z_0, R_0$) are reported in Table~\ref{table:param}. The coordinate $R_0$ is determined by the condition that the blob is located near the boundary radius of the jet. In the scenario with multiple blobs ($N_{\rm blobs}=10$), the blobs are spaced over the jet at the initial time $t=0$. The initial coordinates of the first blob are assigned as in the scenario with a single blob. The initial coordinates of the other blobs are drawn from a uniform distribution in the range $0<\phi<2\pi$, $z_0<z<1.5\;z_0$ (where $z_0$ is the initial distance of the first blob from the SMBH). In the axisymmetric jet (which is modeled as an ensemble of $N_{\rm jet}=10^5$ blobs), the initial coordinates of the blobs are drawn from the same uniform distribution in the range $0<\phi<2\pi$, $z_0<z<1.5\;z_0$. We use a large simulation volume so that the blobs do not leave the volume on the timescale of our simulation as they move along the jet.

The power-law index of the electrons is $p=4$. The corresponding spectrum of synchrotron radiation is $F_\nu \,\propto \,\nu^{-\alpha}$, where $\alpha=(p-1)/2=1.5$ is the photon index. The X-ray spectral index of HSP blazars is typically $\alpha \sim 1-2$, which is similar to the optical spectral index of LSP and ISP blazars \citep[e.g.,][]{fossati98, Abdo2011, ghisellini17}. The magnitude of the polarization degree can change by a factor of a few for different values of $p$. However, the temporal pattern remains qualitatively similar.

\renewcommand{\arraystretch}{1.5}
\renewcommand{\arraystretch}{1.5}
\begin{table*}
\caption{\label{table:param} Initial and final values of the parameters in the scenario with a single blob.}
\centering
\begin{tabular}{l|cccc|ccc|cc}  
\toprule
Jet shape & $\phi_0\, (\rm rad)$ & $z_0\, (\rm cm)$ & $R_0\, (\rm cm)$ & $\Gamma_0$ &  $z_{\rm f}\, (\rm cm)$ & $R_{\rm f} \, (\rm cm)$ &$\Gamma_{\rm f}$  & $\Gamma_{\rm min}$ & $\Gamma_{\rm max}$ \\
\midrule
Cylindrical & $1.92\, \pi$    & $3.91\times 10^{17}$  & $2.98\times 10^{16}$  &10  & $3.73\times 10^{18}$   & $2.98\times 10^{16}$ & 10 & 10 & 10 \\
\midrule
\shortstack{``Sausage-like''\\ $C_2=0.8$}   
  & \raisebox{0.6ex}{$1.40\, \pi$}    
  & \raisebox{0.6ex}{$5.77\times 10^{17}$}  
  & \raisebox{0.6ex}{$1.76\times 10^{16}$}  
  & \raisebox{0.6ex}{7}  
  & \raisebox{0.6ex}{$3.41\times 10^{18}$}   
  & \raisebox{0.6ex}{$3.46\times 10^{16}$}  
  & \raisebox{0.6ex}{10.26}  
  & \raisebox{0.6ex}{6.20}  
  & \raisebox{0.6ex}{11.61} \\ [1.2ex]
\shortstack{``Sausage-like''\\ $C_2=2$}  
  & \raisebox{0.6ex}{$1.45\, \pi$}    
  & \raisebox{0.6ex}{$3.79\times 10^{17}$}  
  & \raisebox{0.6ex}{$1.36\times 10^{16}$}  
  & \raisebox{0.6ex}{7}   
  & \raisebox{0.6ex}{$2.54\times 10^{18}$}   
  & \raisebox{0.6ex}{$2.97\times 10^{16}$} 
  & \raisebox{0.6ex}{10.03} 
  & \raisebox{0.6ex}{4.80} 
  & \raisebox{0.6ex}{12.65} \\
\midrule
Nearly parabolic A & $0.015\, \pi$    & $2.02\times 10^{17}$  & $2.12\times 10^{16}$  &7   & $7.21\times 10^{18}$   & $8.89\times 10^{16}$ & 29.48 & 7 & 29.48 \\
Nearly parabolic B & $1.25\, \pi$    & $2.02\times 10^{17}$  & $2.12\times 10^{16}$  &7   & $5.69\times 10^{18}$   & $8.08\times 10^{16}$ & 26.81 & 7 & 26.81 \\
\bottomrule
\end{tabular}
\tablefoot{
We report the initial and final values of the blob coordinates $(\phi, z, R)$ and of the bulk Lorentz factor, $\Gamma$, measured at $t_{\rm obs} = 0 {\rm\; days}$ (subscript ``0'') and $t_{\rm obs} = 20 {\rm\; days}$ (subscript ``f''), for different jet shapes. The last two columns give the minimum and maximum values of $\Gamma$ reached along the blob’s trajectory. When the jet shape is nearly parabolic, we consider two cases (hereafter denoted as ``single blob A'' and ``single blob B'') that differ only for the initial coordinates of the blob.
}
\end{table*}

\section{Results}
\label{sec:res}

Very long baseline interferometry imaging of radio emission from AGNs jets suggests that the jet shape is nearly parabolic \citep[][]{Mertens2016, Pushkarev2017, Kovalev2020, Boccardi2021}. Imaging of the jet in M87 suggests that the jet shape is ``sausage-like'', namely, the jet radius oscillates about the parabolic profile \citep[see Fig.~6 of][]{Mertens2016}. However, most observations constrain the jet shape on the scale of tens of parsecs, which is significantly longer than the distance from the SMBH where the blazar emission is produced. In the following, we consider three different jet structures:
\begin{itemize}
\item Cylindrical jet. We set $q=0$, $C_1=2/\pi$, and $C_2=0$ in Eq.~\eqref{eq:Y}. The transverse radius of the jet is constant.
\item ``Sausage-like'' jet. We set $q=0.3$ and $C_1 = \left(2-4q \right)/\pi$ in Eq.~\eqref{eq:Y}. We consider two different cases: $C_2 = 0.8$ and $C_2 = 2$. The transverse radius of the jet oscillates about the equilibrium position, which is $R\,\propto\, z^q$. The amplitude of the oscillations increases for large $C_2$ (there would be no oscillations for $C_2=0$).
\item Nearly parabolic jet. We set $q=0.4$, $C_1 = \left(2-4q \right)/\pi$, and $C_2=0$ in Eq.~\eqref{eq:Y}. The transverse radius is $ R\,\propto\, z^q$.
\end{itemize}
In Fig.~\ref{fig:RandGamma} we show the transverse radius and the bulk Lorentz factor of the jet as a function of the distance to the black hole for different jet structures. For each jet structure, we calculate the time-dependent Stokes parameters for the emission scenarios described in Sect.~\ref{sec:timepol}. Our results are summarized in Table~\ref{table:results}.

For the cylindrical jet structure ($q=0$), the temporal profiles of the Stokes parameters of the single blob are periodic. Different initial conditions of the blob would produce a phase shift in the observed emission. For the other jet structures ($q\neq 0$), the temporal profiles are not strictly periodic. The initial conditions of the blob can affect the observed emission. To illustrate this effect, for $q=0.4$, we investigate two models (``single blob A'' and ``single blob B'') that differ in the initial phase of the blob. In the scenario with multiple blobs, the initial phases of the blobs are chosen randomly. Since the number of blobs is relatively large ($N_{\rm blobs}=10$), the temporal profiles of the Stokes parameters are qualitatively similar. For example, the amplitude of the EVPA rotations changes by $<50\%$ for different realizations.

\subsection{Cylindrical jet}

\begin{figure*}
    \centering
    \includegraphics[width=0.33\textwidth]{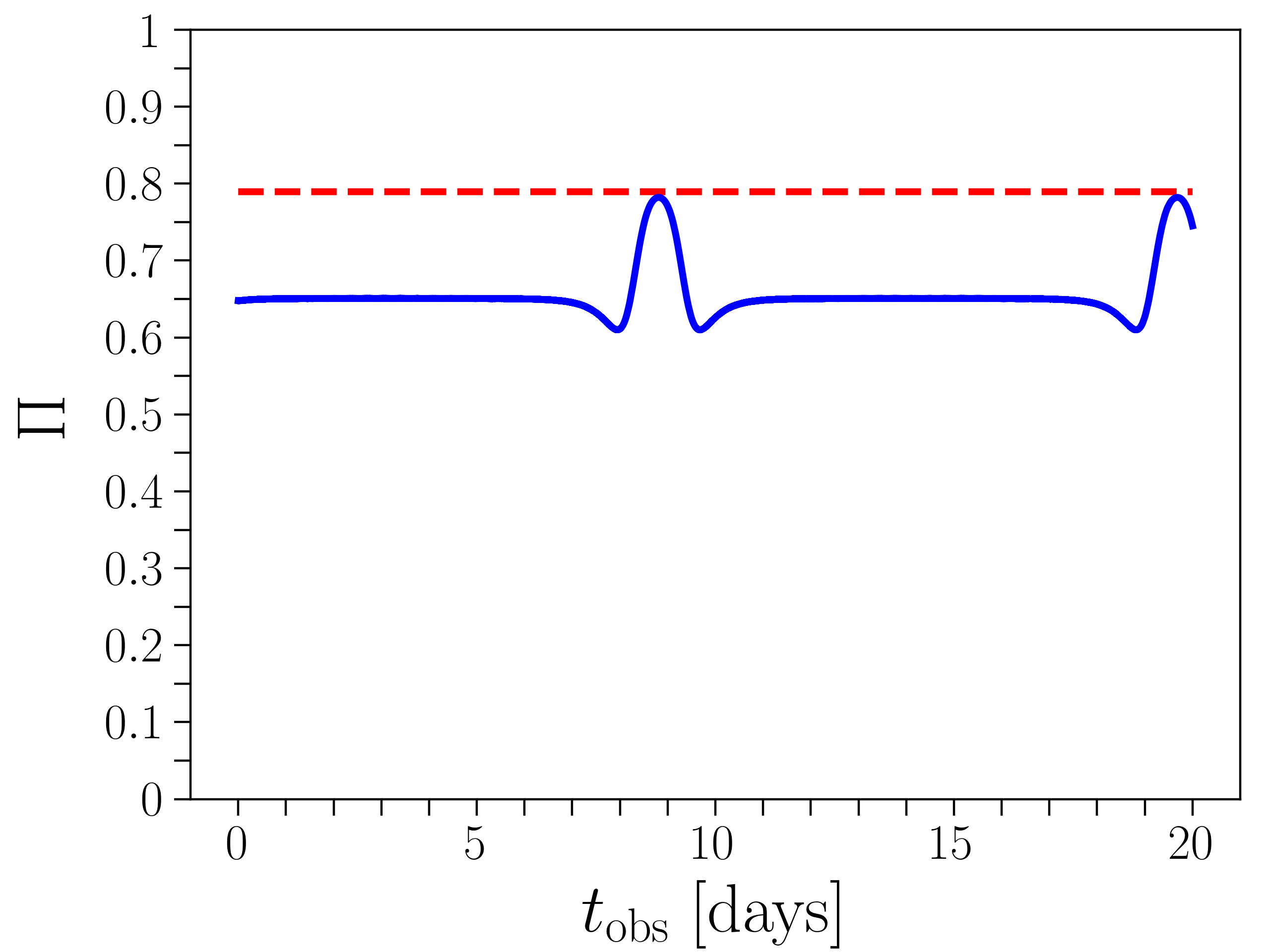}%
    \hspace{+0.25em}%
    \includegraphics[width=0.33\textwidth]{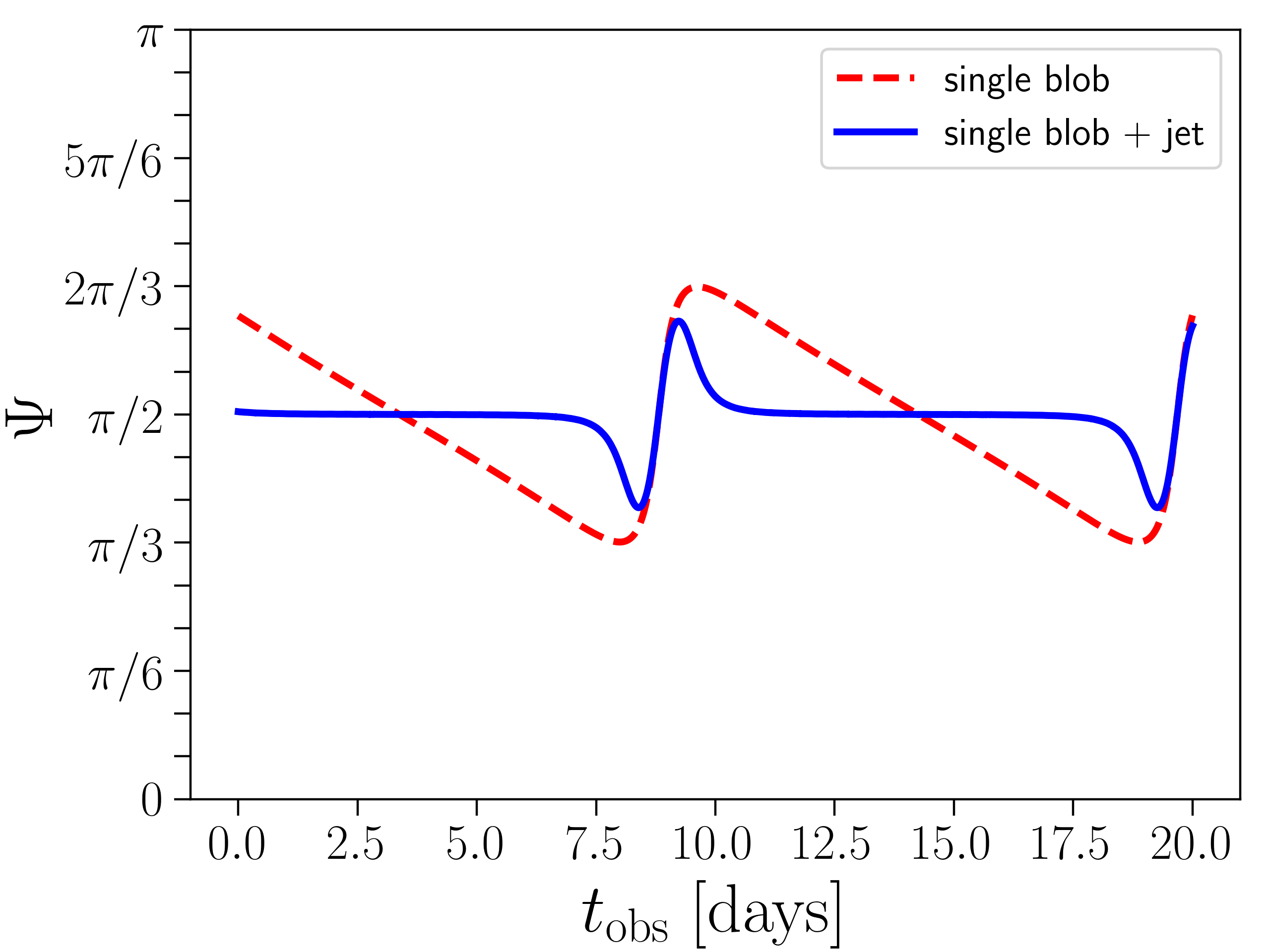}%
    \hspace{+0.25em}%
    \includegraphics[width=0.33\textwidth]{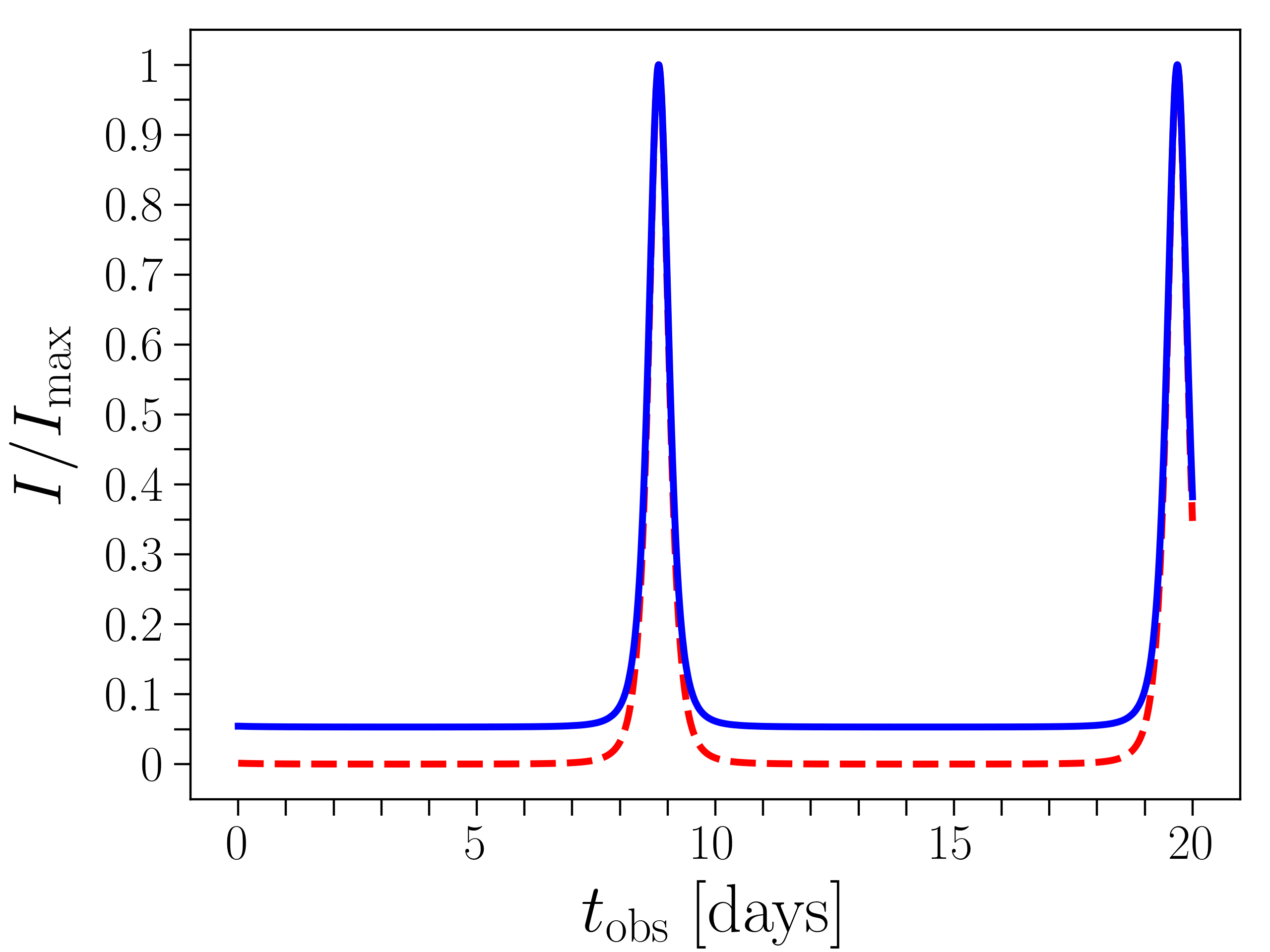}
    \vspace{1mm}
    \includegraphics[width=0.33\textwidth]{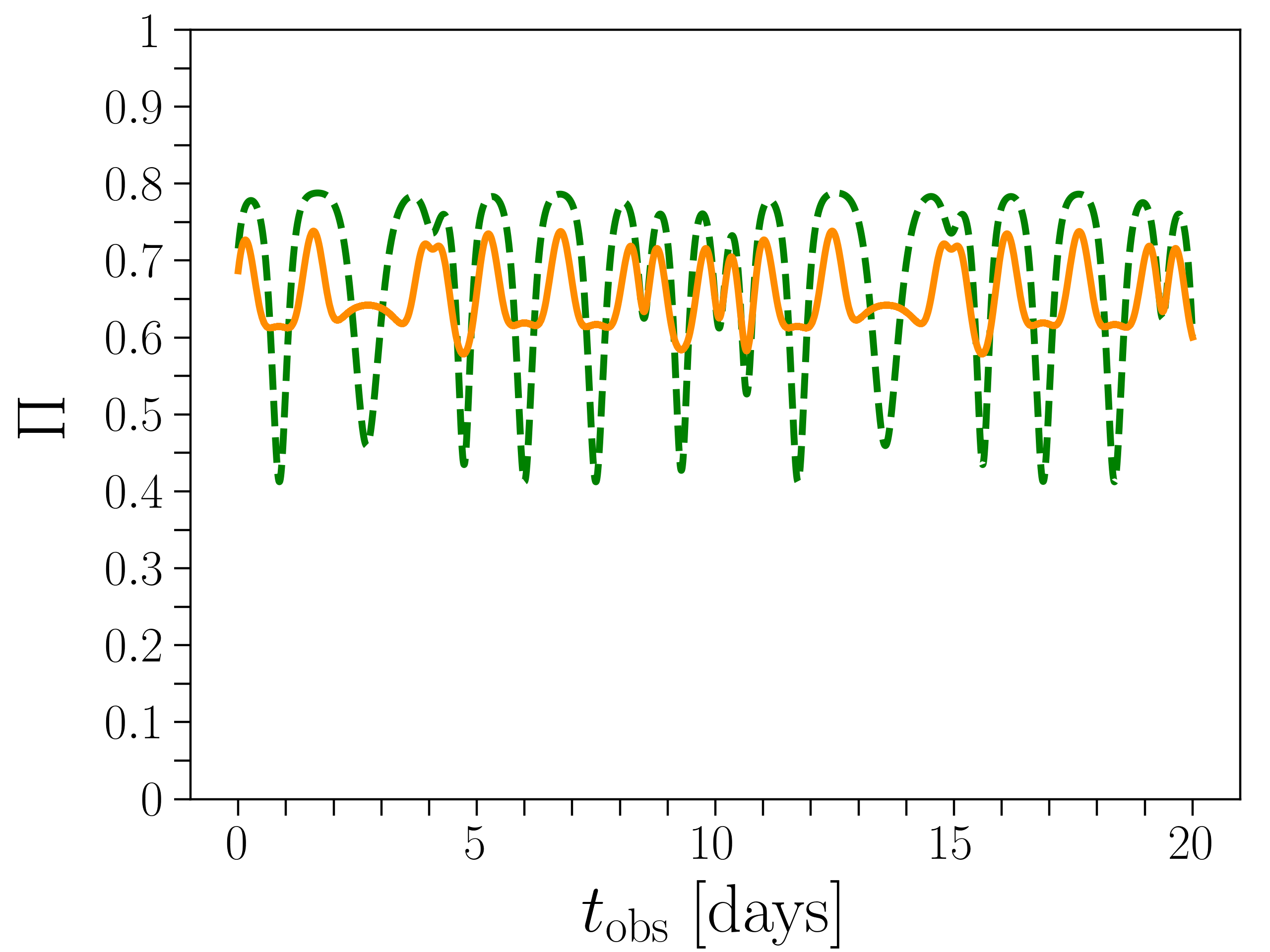}%
    \hspace{+0.25em}%
    \includegraphics[width=0.33\textwidth]{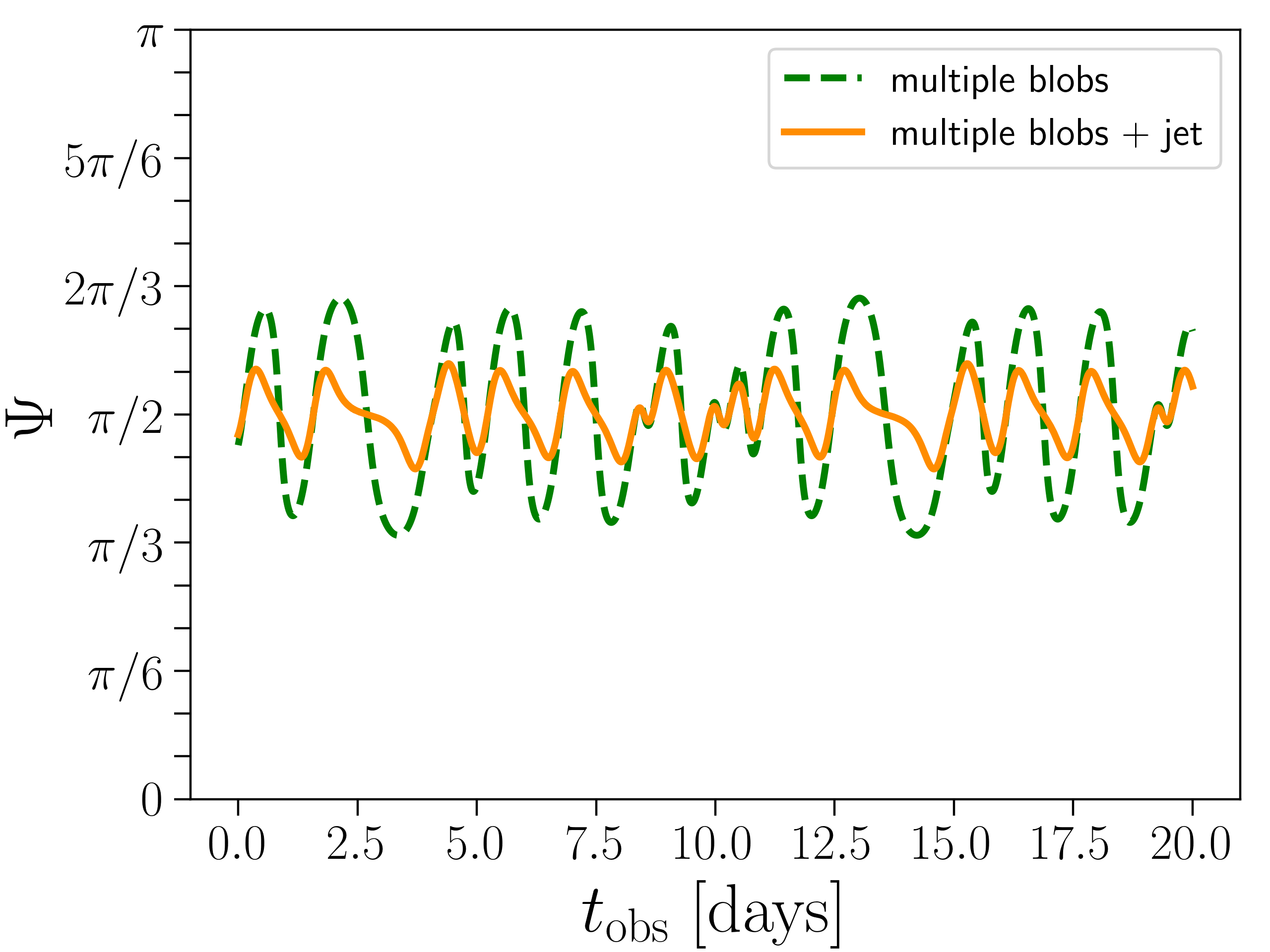}%
    \hspace{+0.25em}%
    \includegraphics[width=0.33\textwidth]{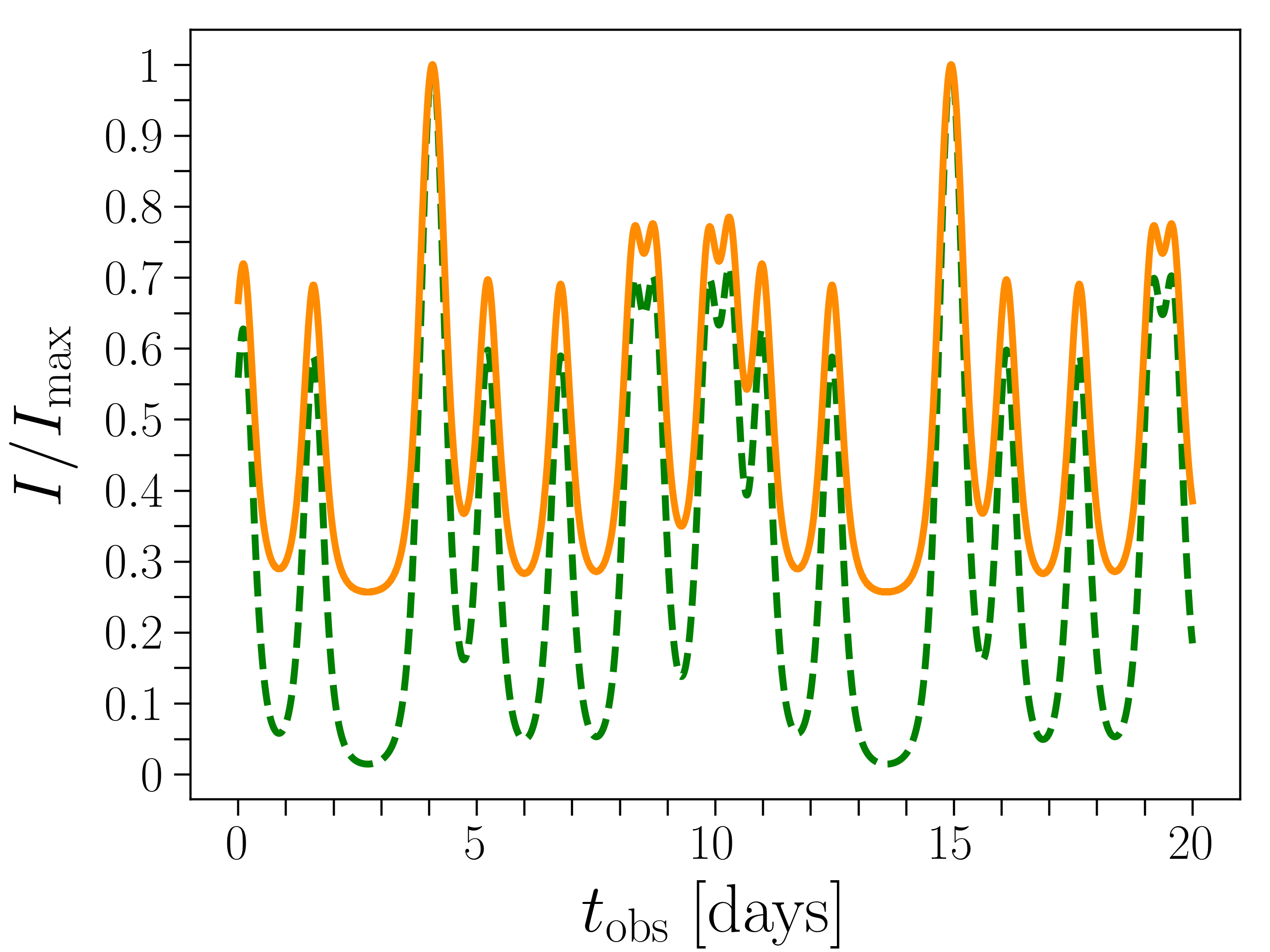}
    \caption{
    Polarization degree $\Pi$ (left panels), electric vector position angle (EVPA) $\Psi$ (middle panels), and intensity normalized to its maximum value $I/I_{\rm max}$ (right panels) as a function of time in the observed frame, $t_{\rm obs}$, measured in days. The  shape of the jet is cylindrical. Top panels: emission from a single blob ($N_{\rm blobs}=1$ and $\eta_{\rm jet}=0$; dashed line) and from a single blob plus a nearly axisymmetric jet ($N_{\rm blobs}=1$ and $\eta_{\rm jet}=1$; solid line). Bottom panels: emission from multiple blobs ($N_{\rm blobs}=10$ and $\eta_{\rm jet}=0$; dashed line) and from multiple blobs plus a nearly axisymmetric jet ($N_{\rm blobs}=10$ and $\eta_{\rm jet}=1$; solid line).
    }
    \label{fig:cilinder}
\end{figure*}

In Fig.~\ref{fig:cilinder}, we show the time evolution of the polarization degree $\Pi$ (left panels), the EVPA $\Psi$ (middle panels), and the intensity normalized to its maximum value $I/I_{\rm max}$ (right panels) for a cylindrical jet. In the top panels, we consider the scenario with a single blob ($N_{\rm blobs}=1$). The nearly axisymmetric jet may or may not contribute to the observed emission (solid and dashed lines correspond, respectively, to $\eta_{\rm jet}=1$ and $\eta_{\rm jet}=0$).

When $\eta_{\rm jet}=0$, the polarization degree is equal to $\Pi_0=(p+1)/(p+7/3)$, as expected for a uniform electromagnetic field (when $N_{\rm blobs}=1$ and $\eta_{\rm jet}=0$, we always have $\Pi=\Pi_0$, independent of the jet shape). The EVPA shows a rapid rotation  $\Delta \Psi \sim 60^{\circ}$ on a timescale of $\Delta t_{\rm obs} \sim 1.5\;\mathrm{days}$, which coincides with the peak of the observed intensity (the peak occurs when the velocity of the blob is aligned with the line of sight). When $\eta_{\rm jet}=1$, the polarization degree is lower than $\Pi_0$ most of the time and approaches $\Pi_0$ when the blob dominates the observed emission. The EVPA is constant ($\Psi\simeq\pi/2$) when the jet dominates the observed emission and varies when the blob dominates.

In the bottom panels of Fig.~\ref{fig:cilinder}, we consider the scenario with multiple blobs ($N_{\rm blobs}=10$). We observe a periodic behavior in $\Pi$, $\Psi$, and $I/I_{\rm max}$ with an observed period of approximately $11\;\mathrm{days}$ (the period is the same as in the scenario with a single blob). The amplitude of the variations is greater when $\eta_{\rm jet}=0$ (we find $\Delta \Pi \sim 0.40$ and $\Delta \Psi \sim 55^{\circ}$). Variations are smaller when $\eta_{\rm jet}=1$ because the Stokes parameters of the jet $(I_{\rm jet}, Q_{\rm jet}, U_{\rm jet})$ are nearly constant. For example, since we model the jet as an ensemble of a very large number of blobs ($N_{\rm jet}=10^5$), intensity variations are erased because at any given time there are many blobs directed along the line of sight.

\subsection{``Sausage-like'' jet}

\begin{figure*}
    \centering
    \includegraphics[width=0.33\textwidth]{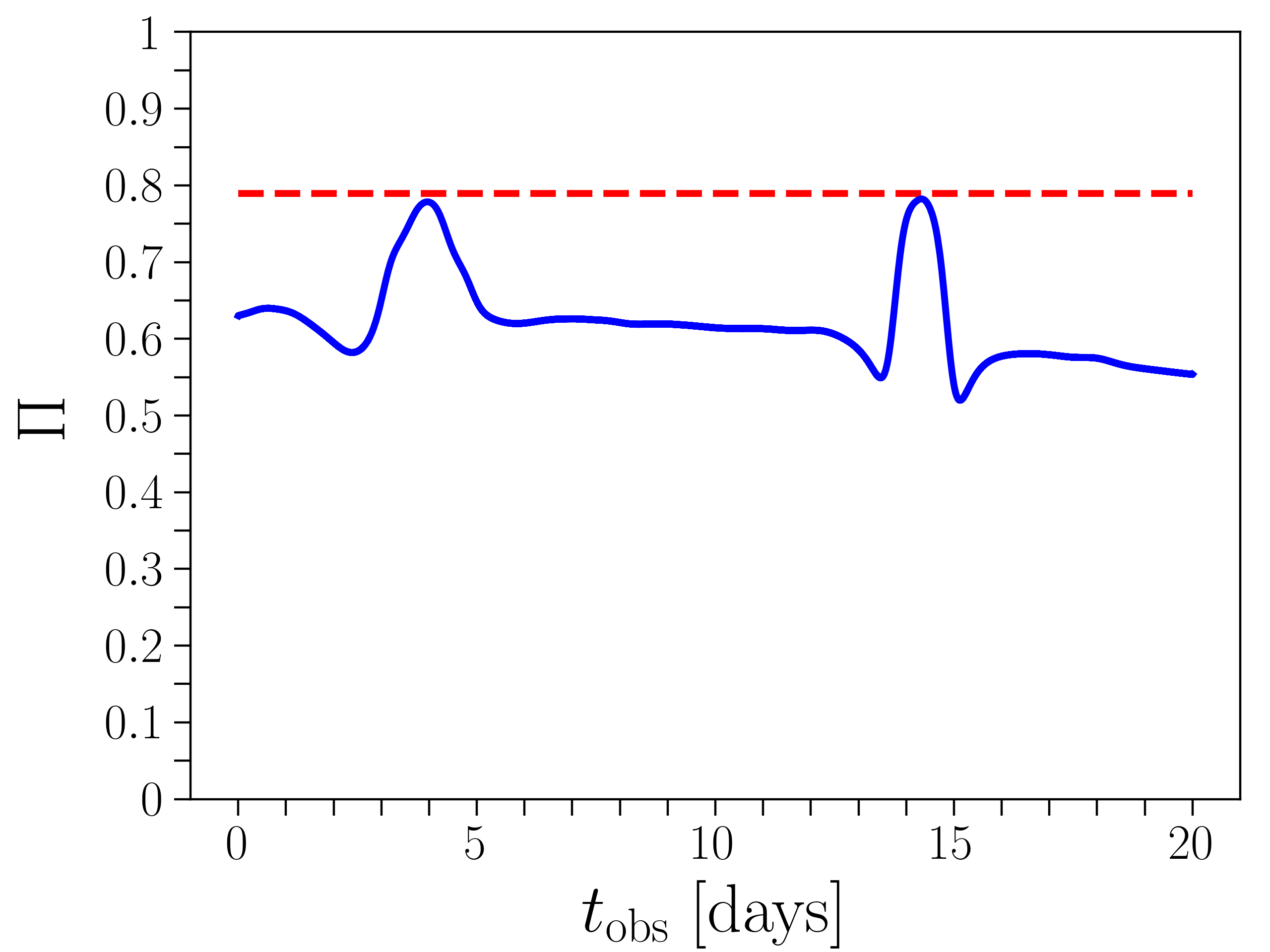}%
    \hspace{+0.25em}%
    \includegraphics[width=0.33\textwidth]{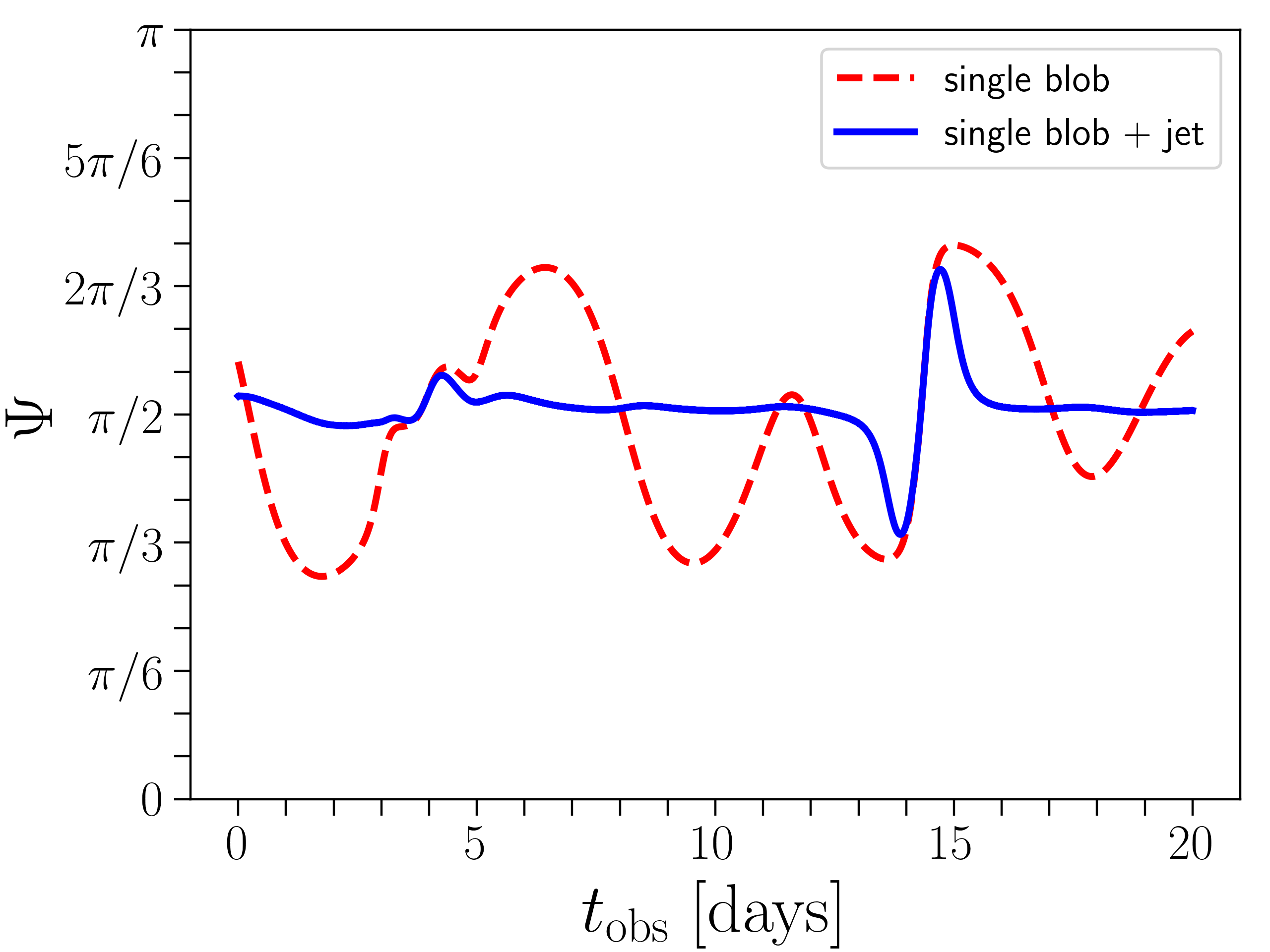}%
    \hspace{+0.25em}%
    \includegraphics[width=0.33\textwidth]{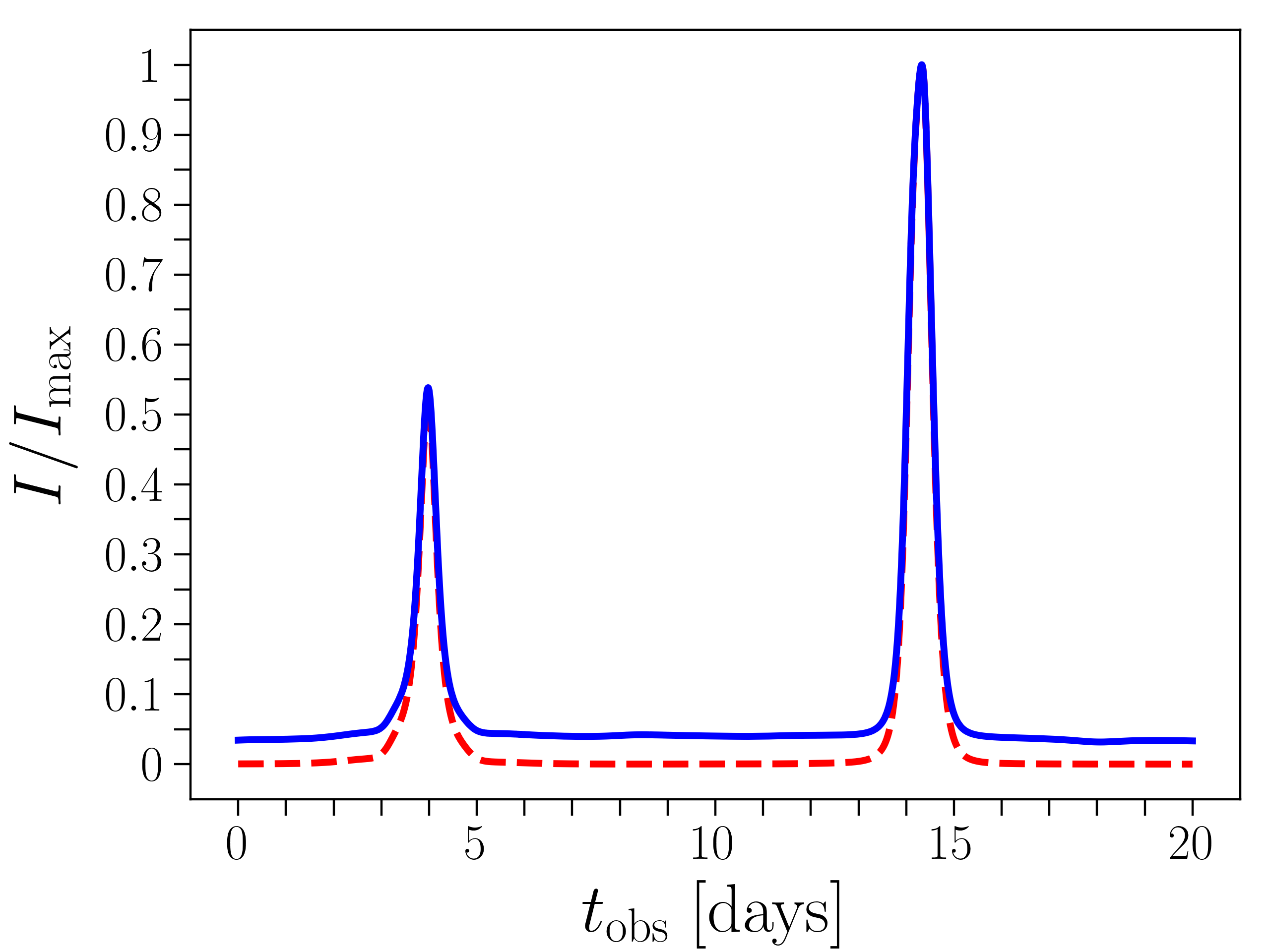}
    \vspace{1mm}
    \includegraphics[width=0.33\textwidth]{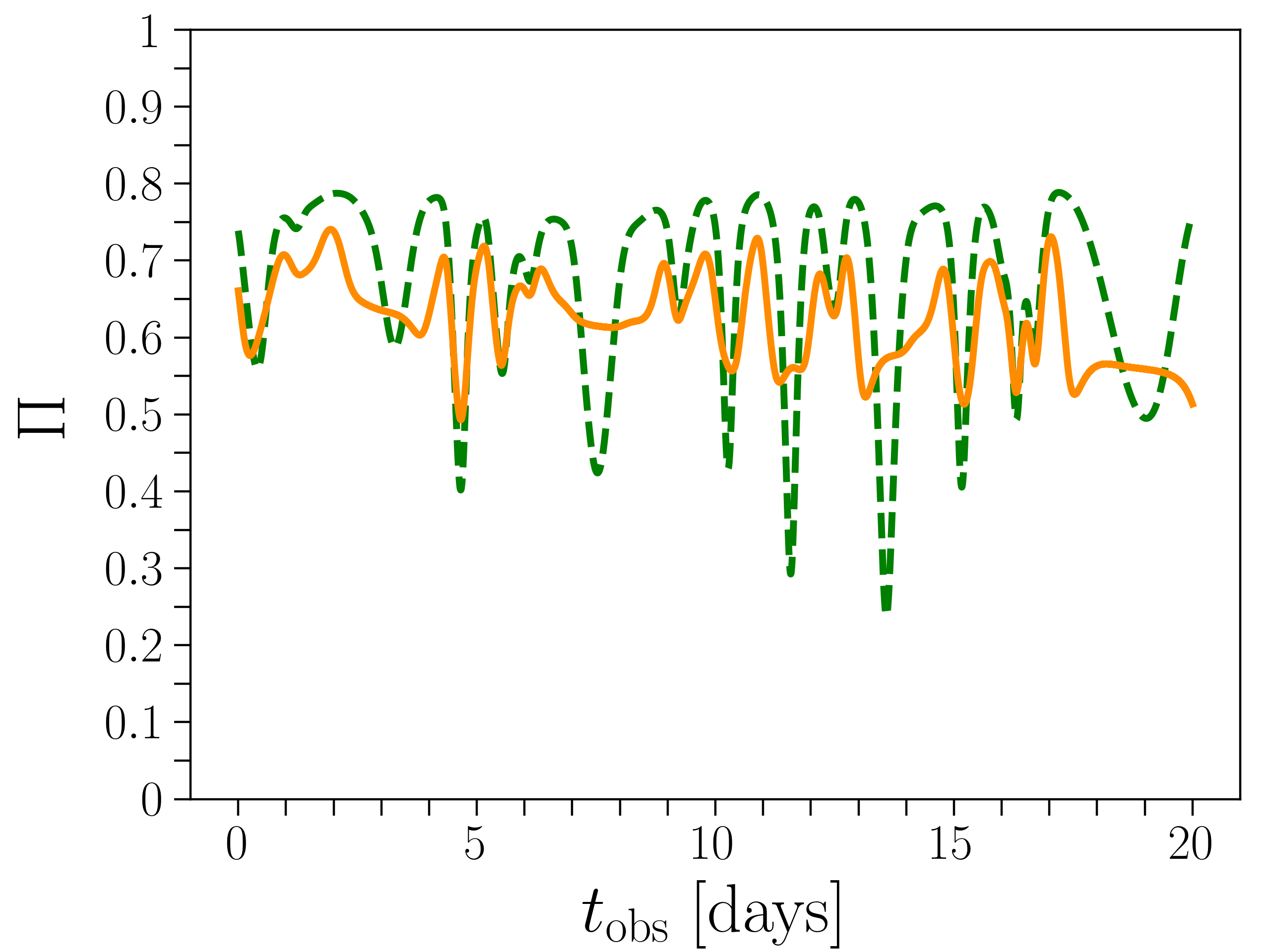}%
    \hspace{+0.25em}%
    \includegraphics[width=0.33\textwidth]{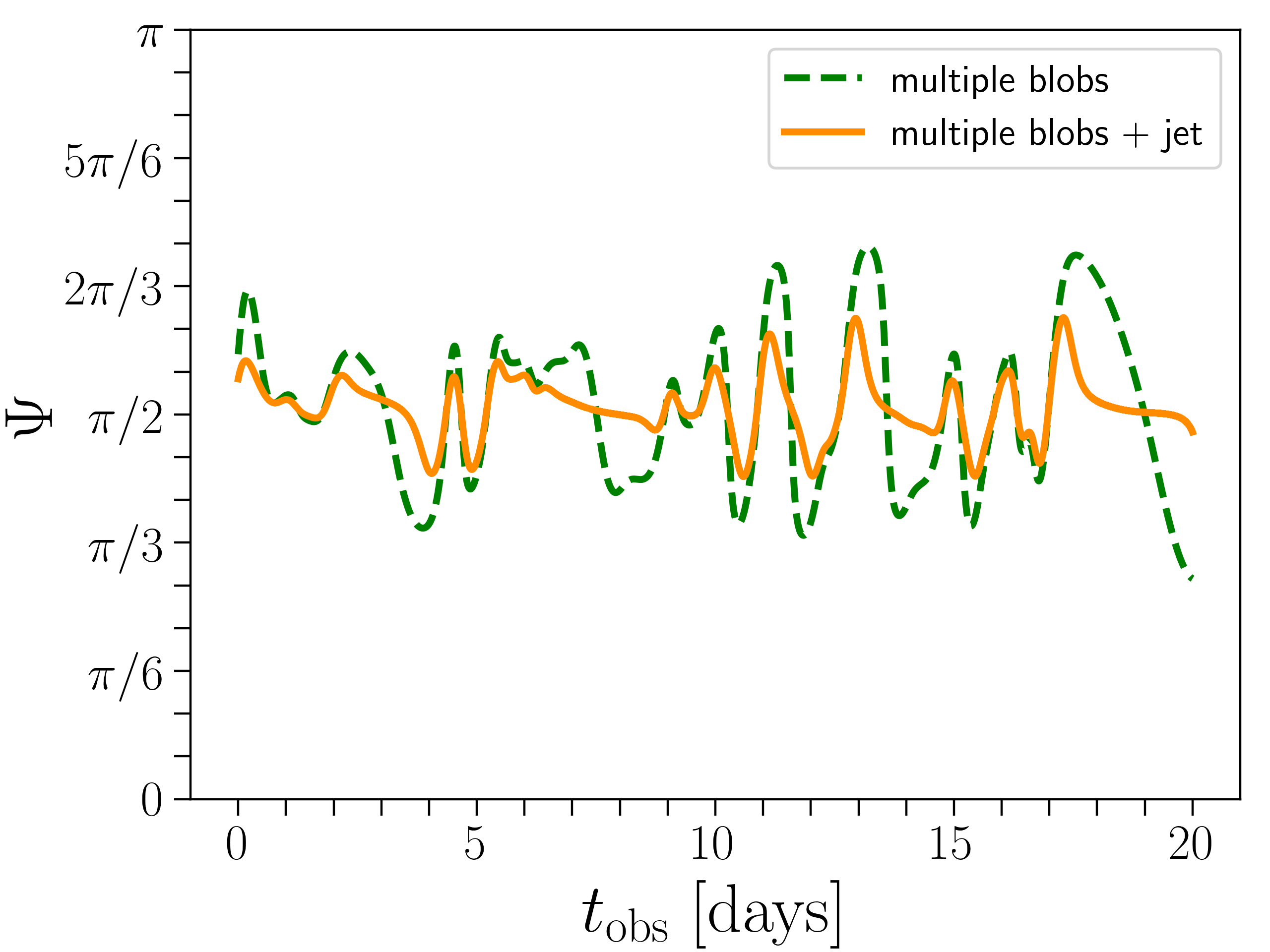}%
    \hspace{+0.25em}%
    \includegraphics[width=0.33\textwidth]{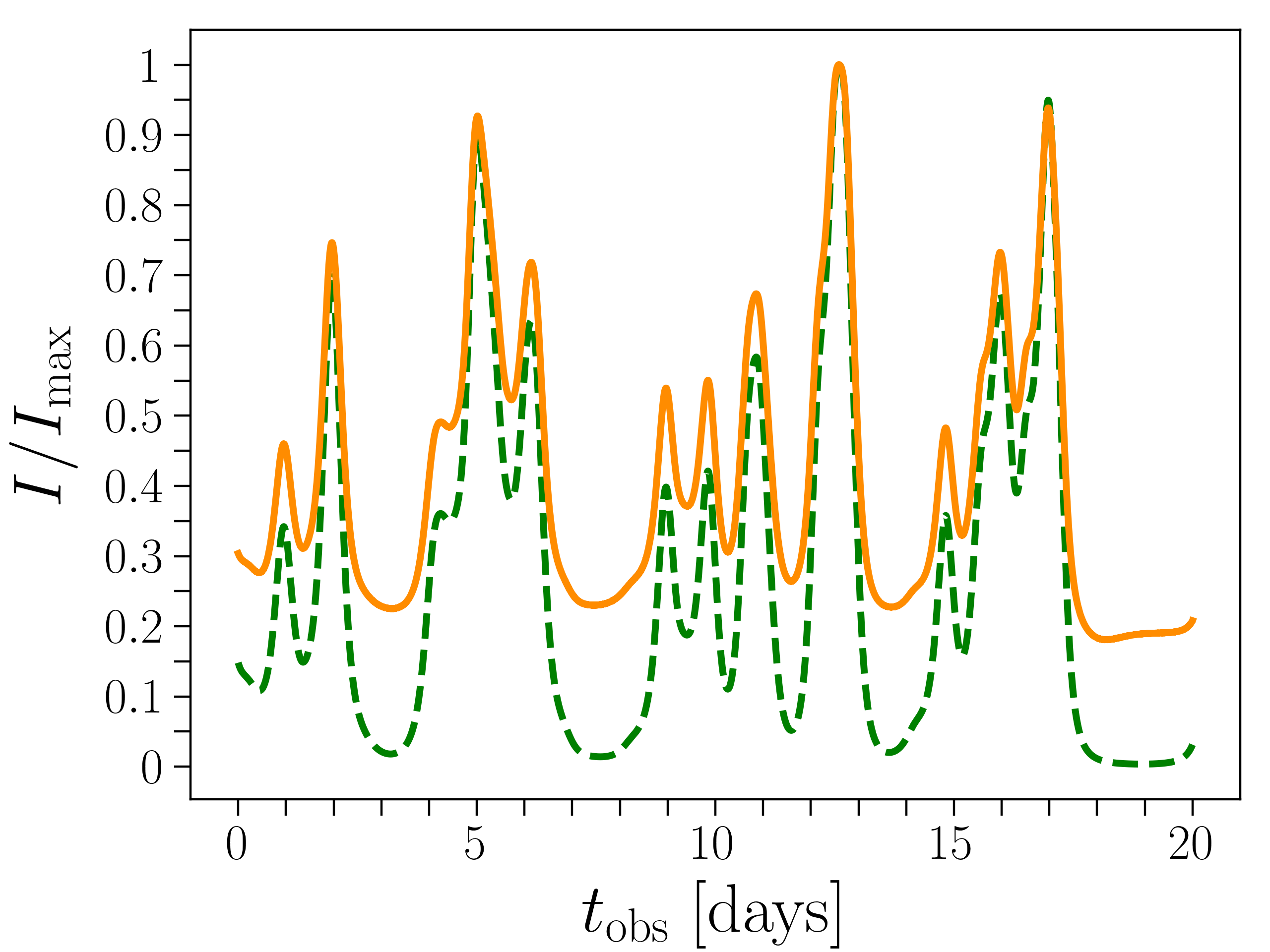}
    \caption{Same as Fig.~\ref{fig:cilinder}, but the shape of the jet is ``sausage-like'' with $C_2=0.8$.}
    \label{fig:sausageA}
\end{figure*}

\begin{figure*}
    \centering
    \includegraphics[width=0.33\textwidth]{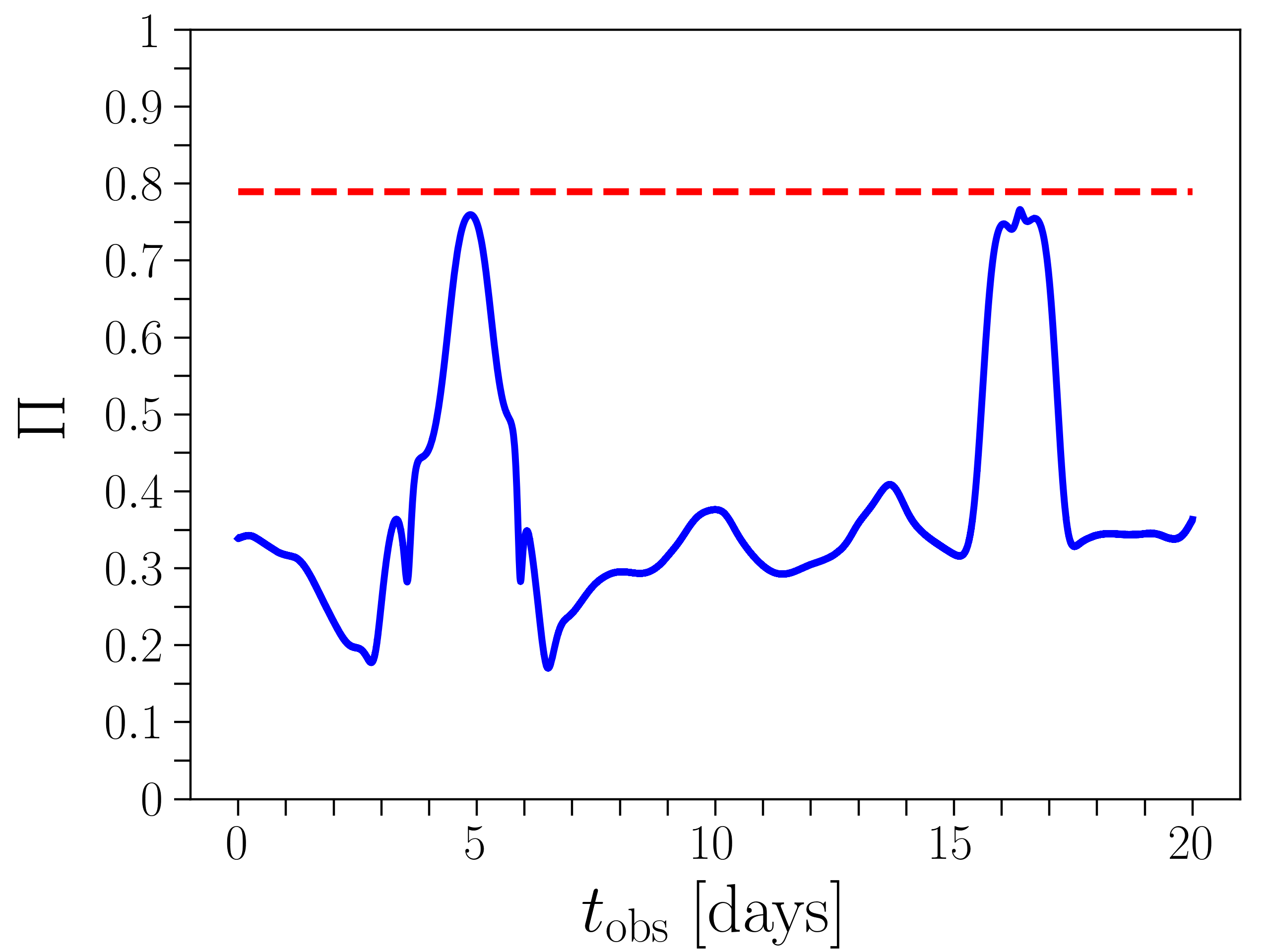}%
    \hspace{+0.25em}%
    \includegraphics[width=0.33\textwidth]{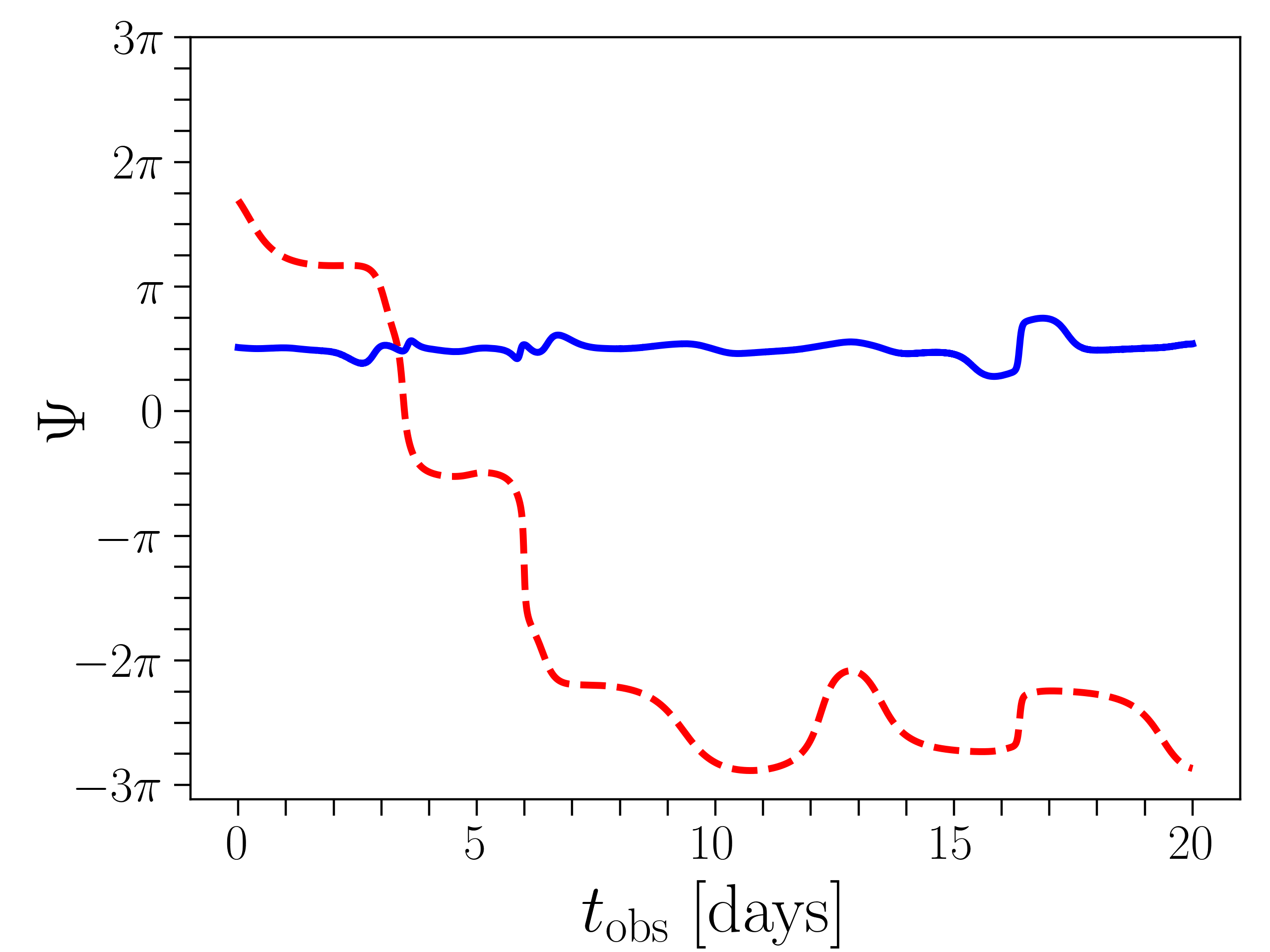}%
    \hspace{+0.25em}%
    \includegraphics[width=0.33\textwidth]{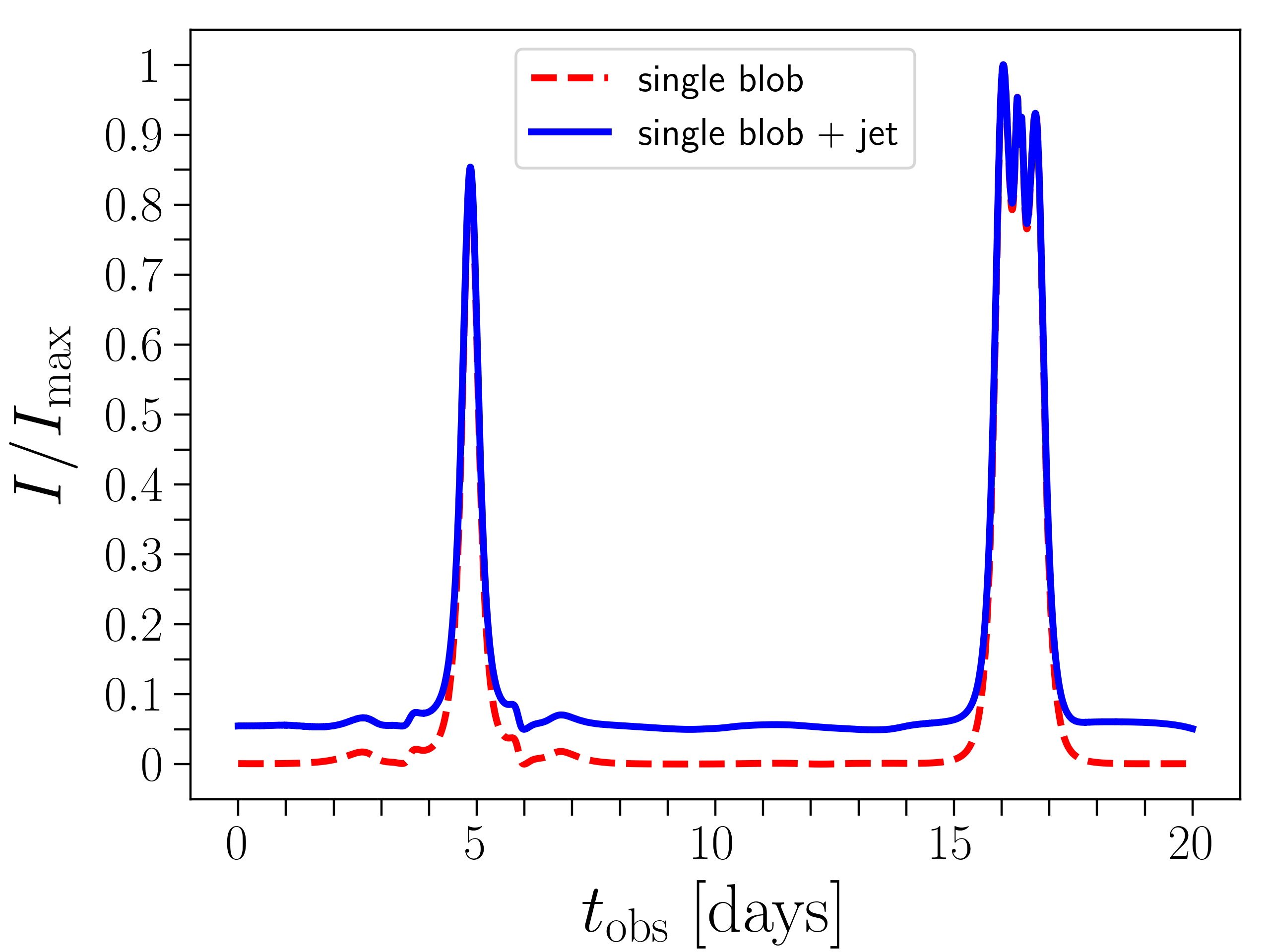}
    \vspace{1mm}
    \includegraphics[width=0.33\textwidth]{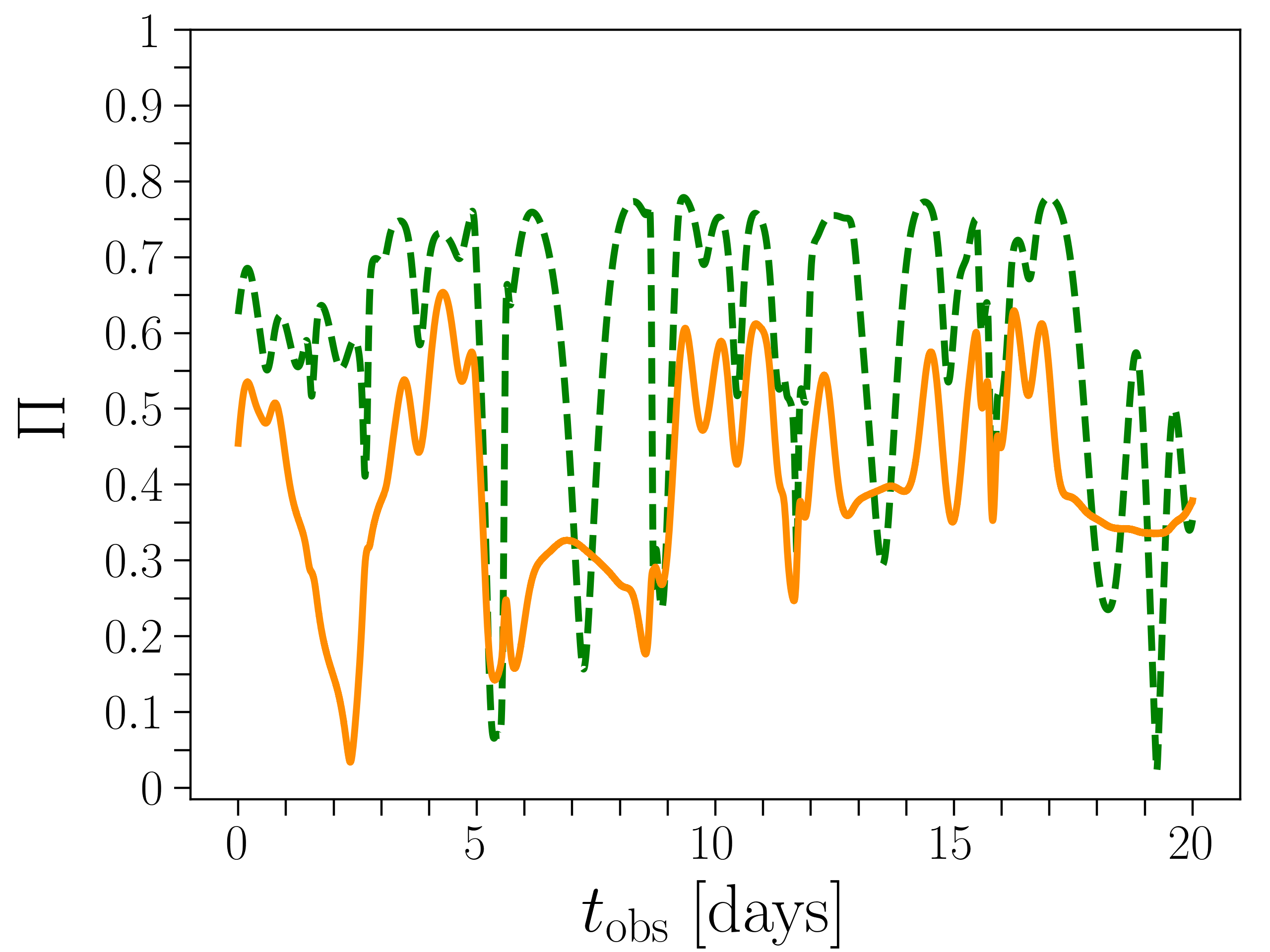}%
    \hspace{+0.25em}%
    \includegraphics[width=0.33\textwidth]{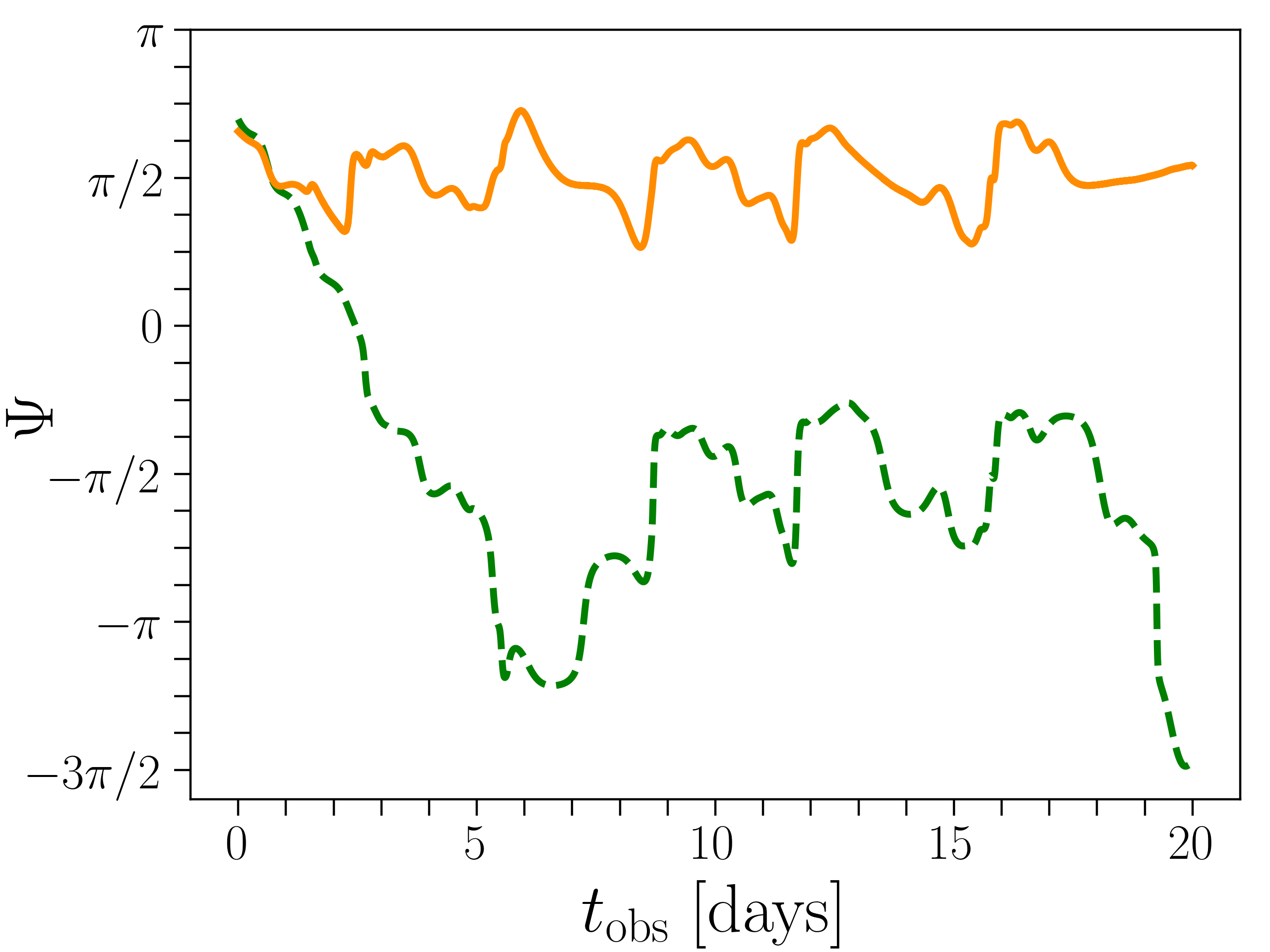}%
    \hspace{+0.25em}%
    \includegraphics[width=0.33\textwidth]{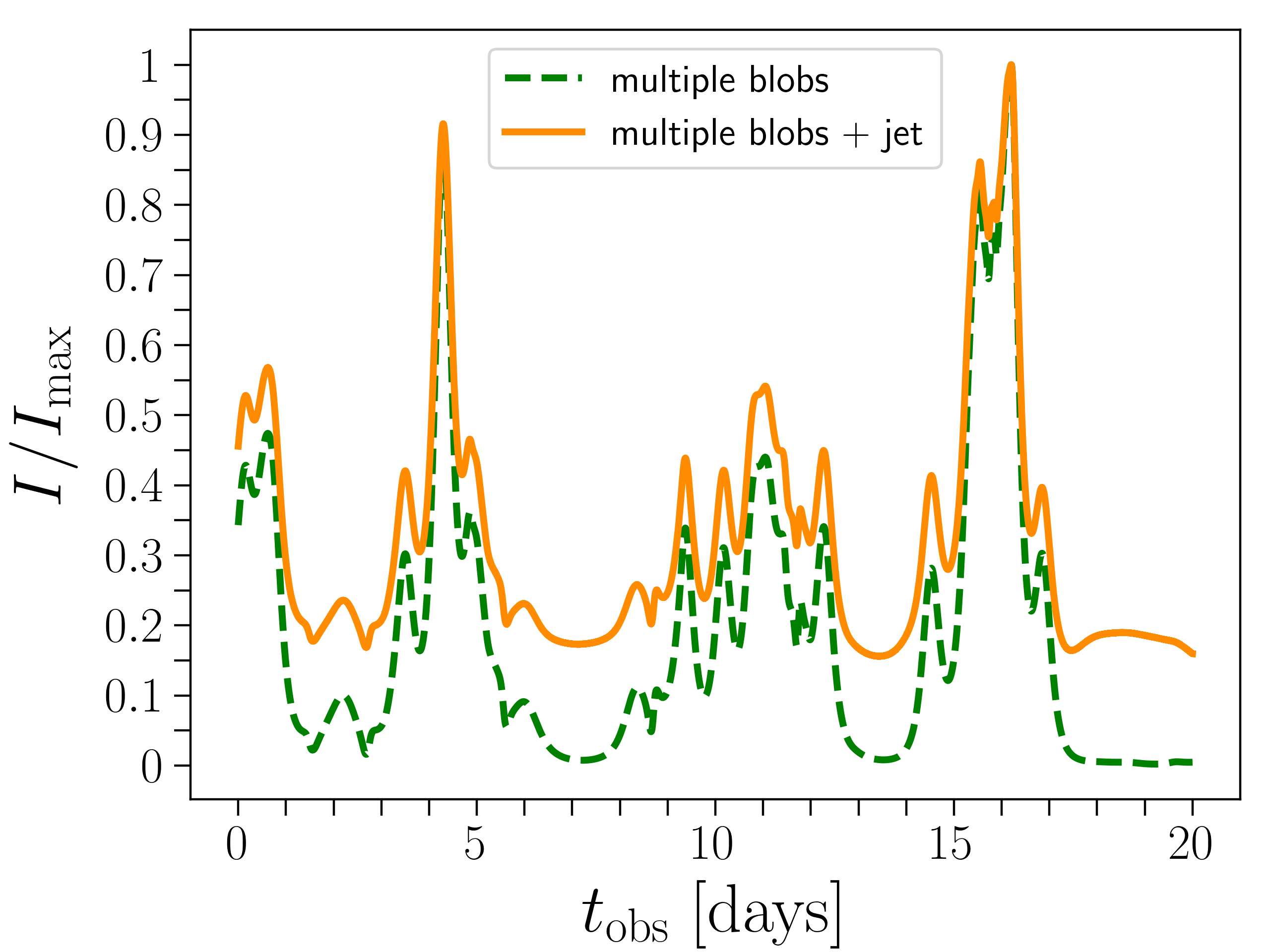}  
    \caption{Same as Fig.~\ref{fig:cilinder}, but the shape of the jet is ``sausage-like'' with $C_2=2$.}
    \label{fig:sausageB}
\end{figure*}

In Figs.~\ref{fig:sausageA} and \ref{fig:sausageB}, we show the time evolution of $\Pi$ (left panels), $\Psi$ (middle panels), and $I/I_{\rm max}$ (right panels) for a ``sausage-like'' jet. In Fig.~\ref{fig:sausageA} we use $C_2 = 0.8$, and in Fig.~\ref{fig:sausageB} we use $C_2 = 2$. The intensity peaks occur when the velocity of a blob is aligned with the line of sight. However, the trajectory of the blobs is complex because the rotation period in the $\phi$ direction is not connected with the oscillations of the jet radius along the $z$ direction. Then, we do not observe a periodic behavior in $\Pi$, $\Psi$, and $I/I_{\rm max}$.

The case $C_2 = 0.8$, which is shown in Fig.~\ref{fig:sausageA}, corresponds to smaller oscillations of the jet transverse radius about the equilibrium position with respect to $C_2 = 2$. In the scenario with a single blob (top panels), when $\eta_{\rm jet}=0$ the EVPA shows smooth variations of amplitude $\Delta \Psi \sim 77^{\circ}$. When $\eta_{\rm jet}=1$, the polarization degree approaches $\Pi_0$ when the blob dominates the observed emission. The EVPA remains nearly constant, except for a sudden jump of approximately $\Delta \Psi \sim 60^{\circ}$ at $t_{\rm obs} \sim 14\;\mathrm{days}$, which occurs when the intensity reaches $I \sim I_{\rm max}$.

In the scenario with multiple blobs (bottom panels), when $\eta_{\rm jet}=0$ we observe significant variability of $\Pi$ and $\Psi$. The variability pattern is complex and seemingly erratic, despite being produced by a deterministic geometric model. We observe a rapid jump in the EVPA of approximately $\Delta \Psi \sim 80^{\circ}$ over a timescale of $1-2$ days at $t_{\rm obs} \sim 12\;\mathrm{days}$, associated with a sharp decrease in the polarization degree ($\Delta \Pi \sim 0.55$). This jump occurs when $I\sim I_{\rm max}$. When $\eta_{\rm jet}=1$, we observe a similar behavior, although the amplitude of the jumps is smaller.

The case $C_2 = 2$, which is shown in Fig.~\ref{fig:sausageB}, corresponds to larger oscillations of the jet transverse radius about the equilibrium position with respect to $C_2 = 0.8$. In the scenario with a single blob (top panels), when $\eta_{\rm jet}=0$ we observe a large EVPA rotation ($\Delta\Psi\sim 720^{\circ})$ over a period of six days. The rotation is associated with an increase in intensity from the minimum to $85\%$ of its maximum value. When the intensity reaches its peak at $t_{\rm obs} \sim 4.5 \; \rm days$ and subsequently decreases, the EVPA undergoes a smaller jump of $\sim 90^{\circ}$. When $\eta_{\rm jet}=1$, the polarization degree and the intensity show two peaks. The polarization degree varies by $\Delta \Pi \sim 0.60$. The EVPA remains nearly constant, except for a sharp rotation of approximately $\Delta \Psi \sim 80^{\circ}$ occurring between $t_{\rm obs} \sim 15$–$16 \;\mathrm{days}$, near the intensity maximum.

In the scenario with multiple blobs (bottom panels), as previously described for a cylindrical jet, we observe a complex evolution of all observables. When $\eta_{\rm jet} = 0$, the polarization degree shows an abrupt jump at $t_{\rm obs} \sim 5 \;\mathrm{days}$, increasing from $\Pi \sim 0$ to $\Pi \sim \Pi_0$ on a timescale of less than a day. During the same time, the EVPA undergoes a rotation of approximately $100^{\circ}$. On a timescale of one week, the EVPA rotates by $\Delta \Psi \sim 390^{\circ}$. When $\eta_{\rm jet}=1$, the variability of the observables is less extreme.

\subsection{Nearly parabolic jet}

\begin{figure*}
    \includegraphics[width=0.33\textwidth]{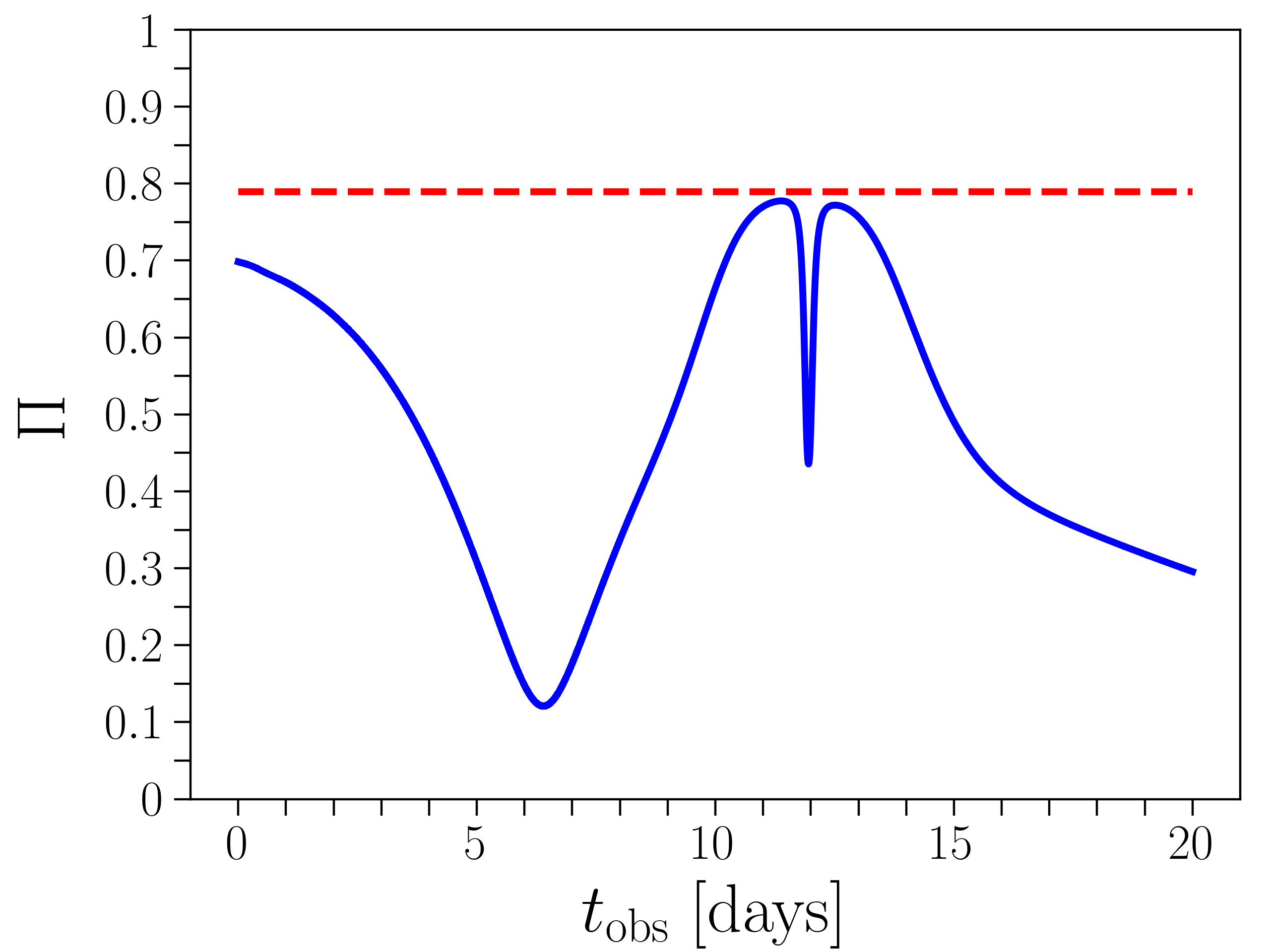}%
    \hspace{+0.25em}%
    \includegraphics[width=0.33\textwidth]{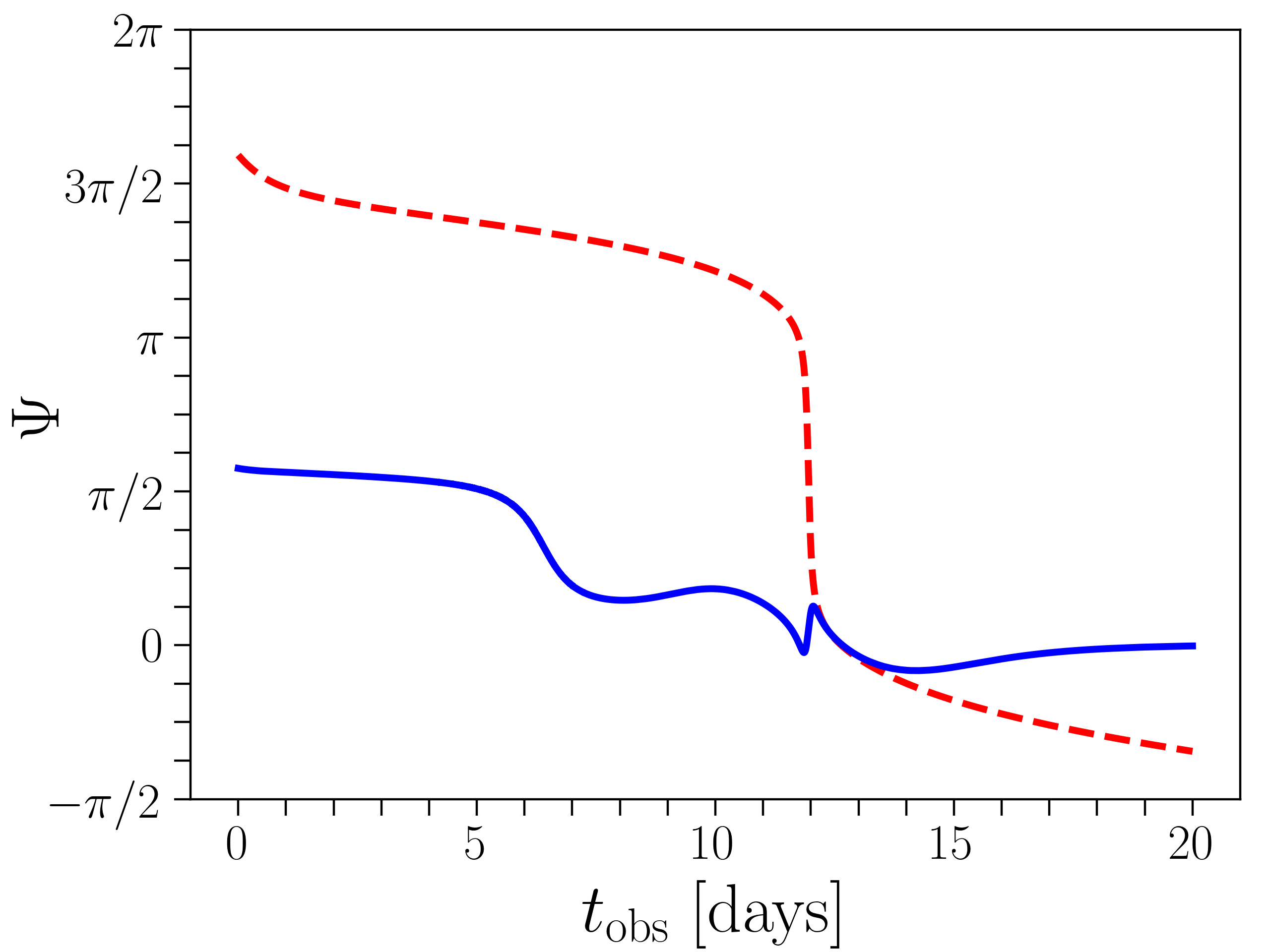}%
    \hspace{+0.25em}%
    \includegraphics[width=0.33\textwidth]{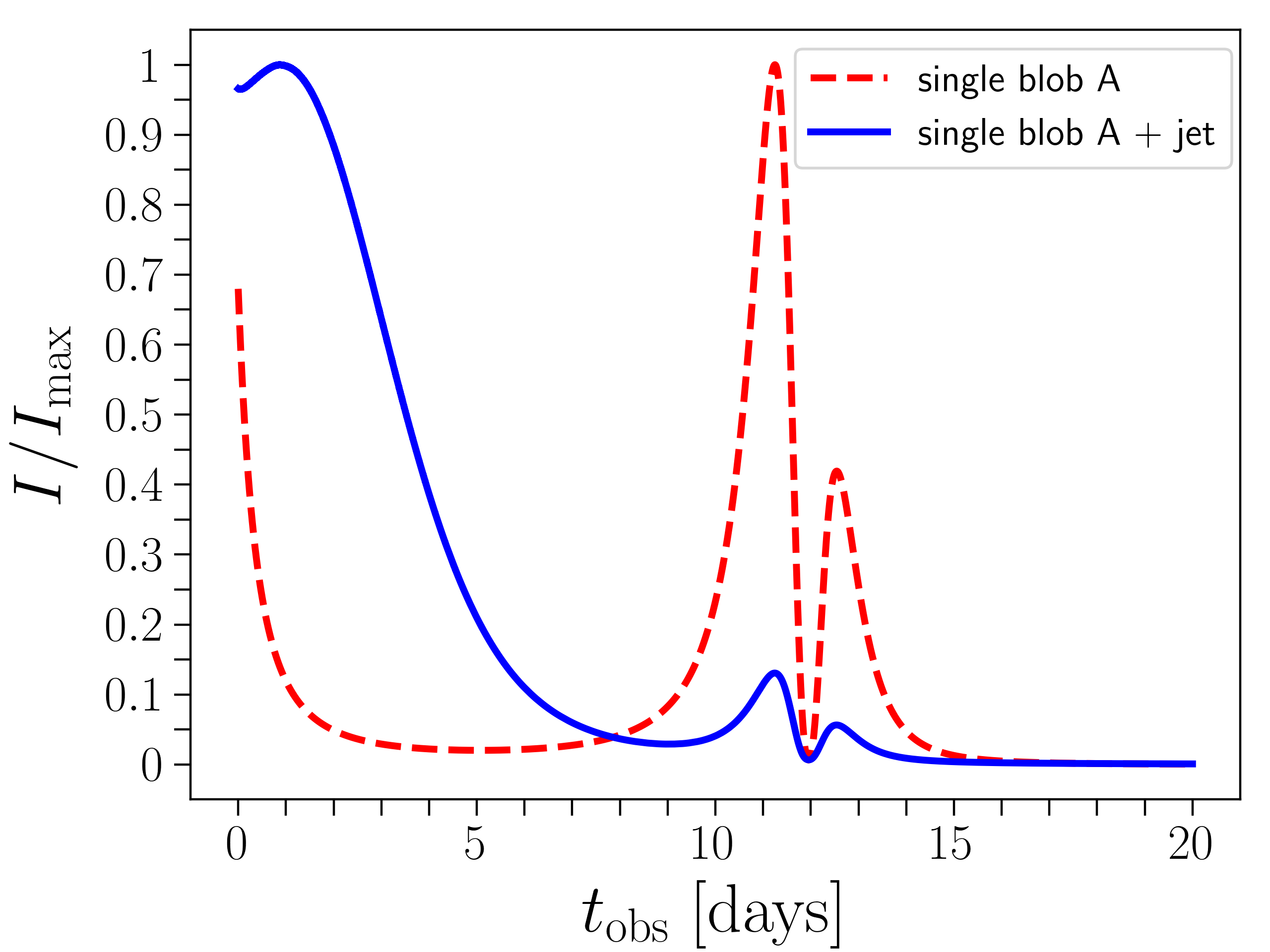}
    \vspace{1mm} 
    \includegraphics[width=0.33\textwidth]{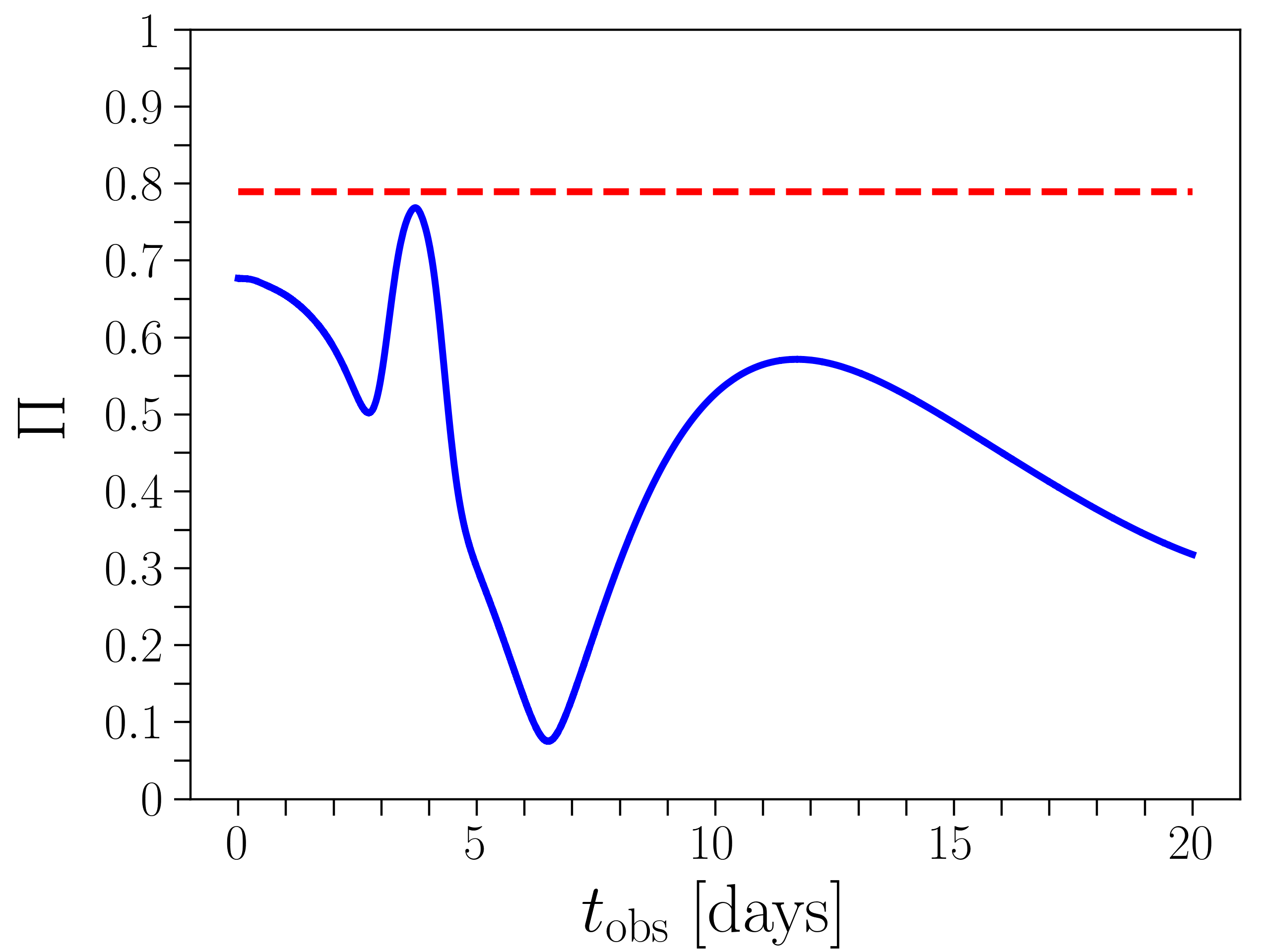}%
    \hspace{+0.25em}%
    \includegraphics[width=0.33\textwidth]{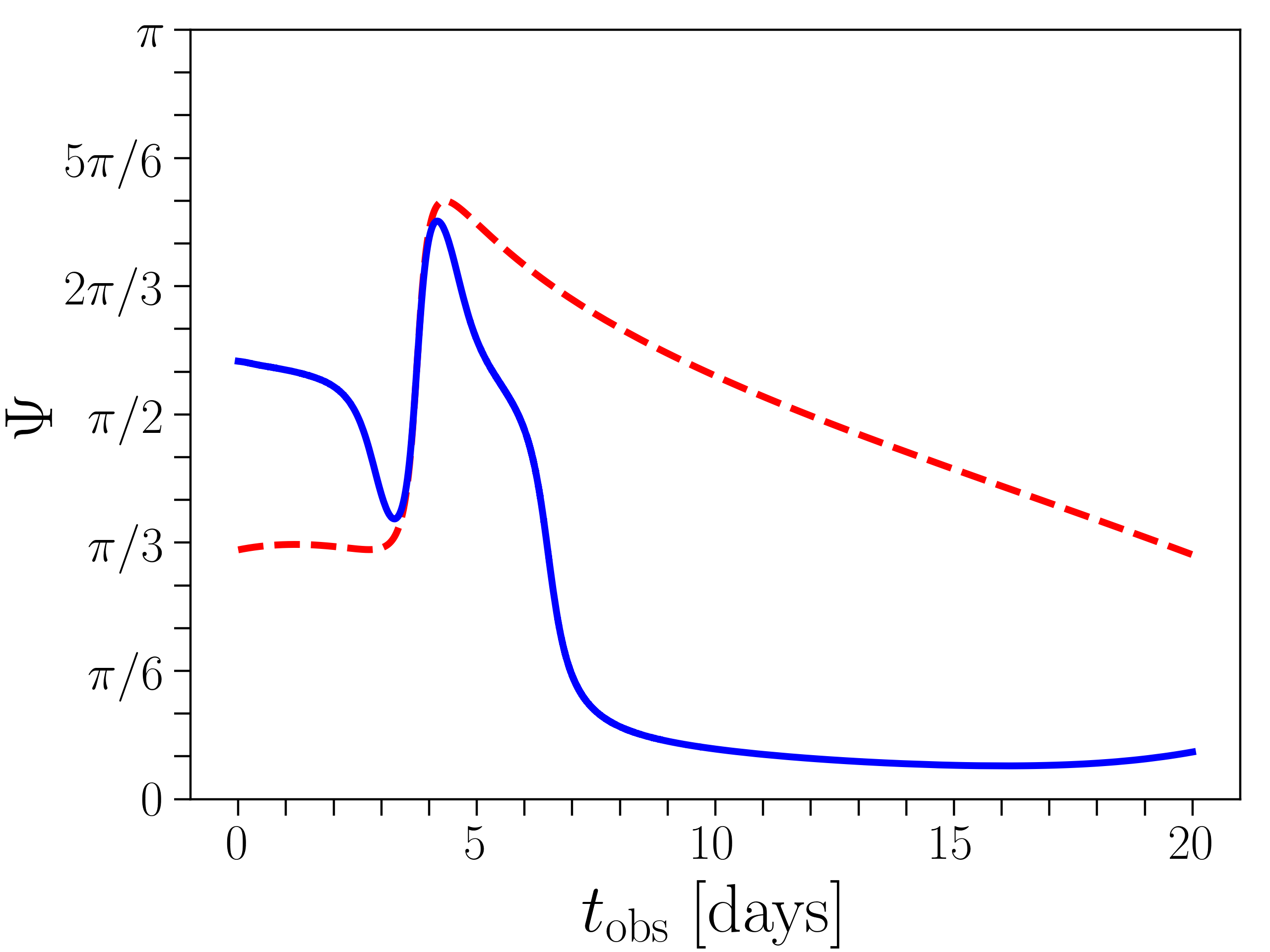}%
    \hspace{+0.25em}%
    \includegraphics[width=0.33\textwidth]{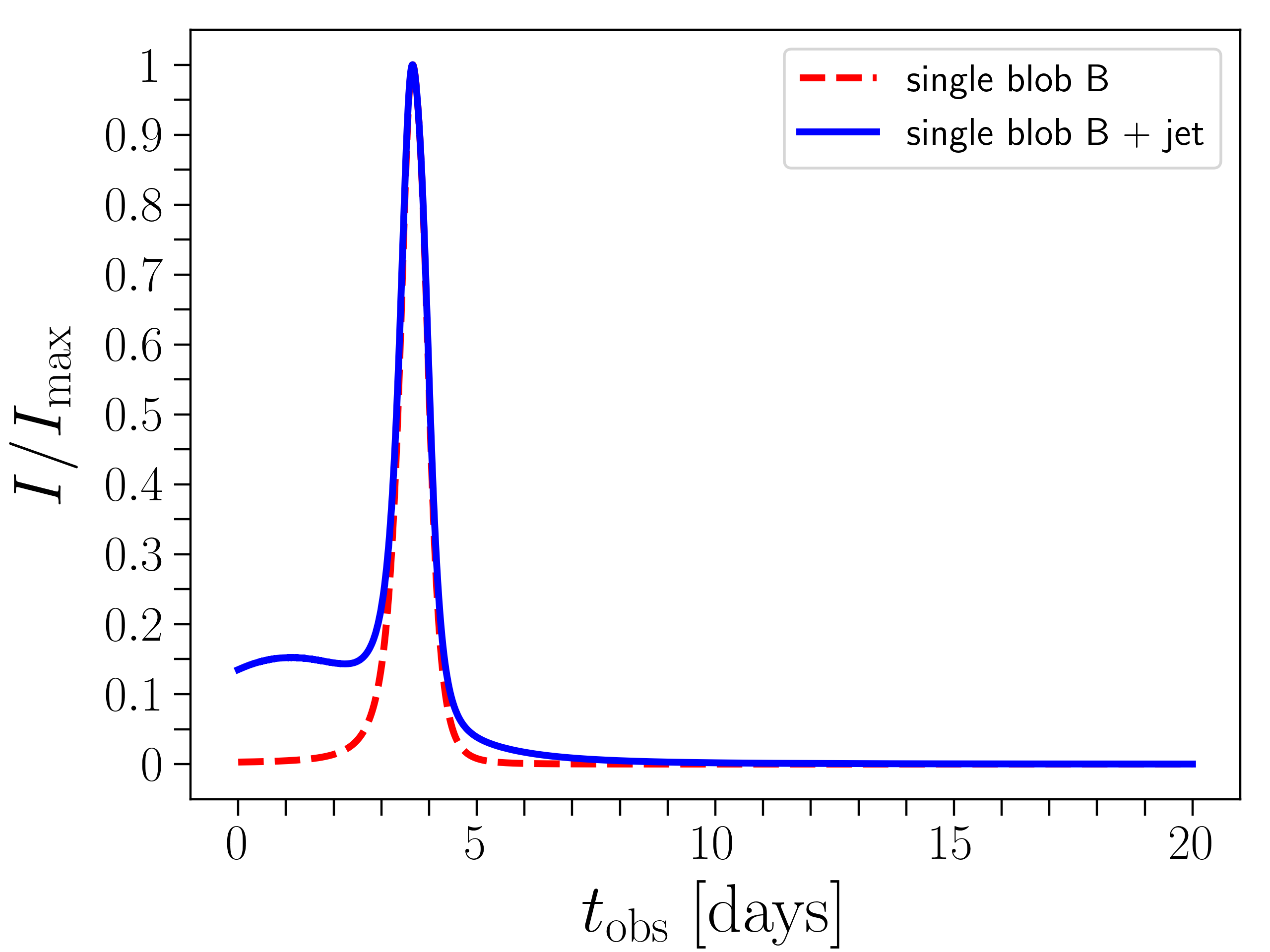}
    \includegraphics[width=0.33\textwidth]{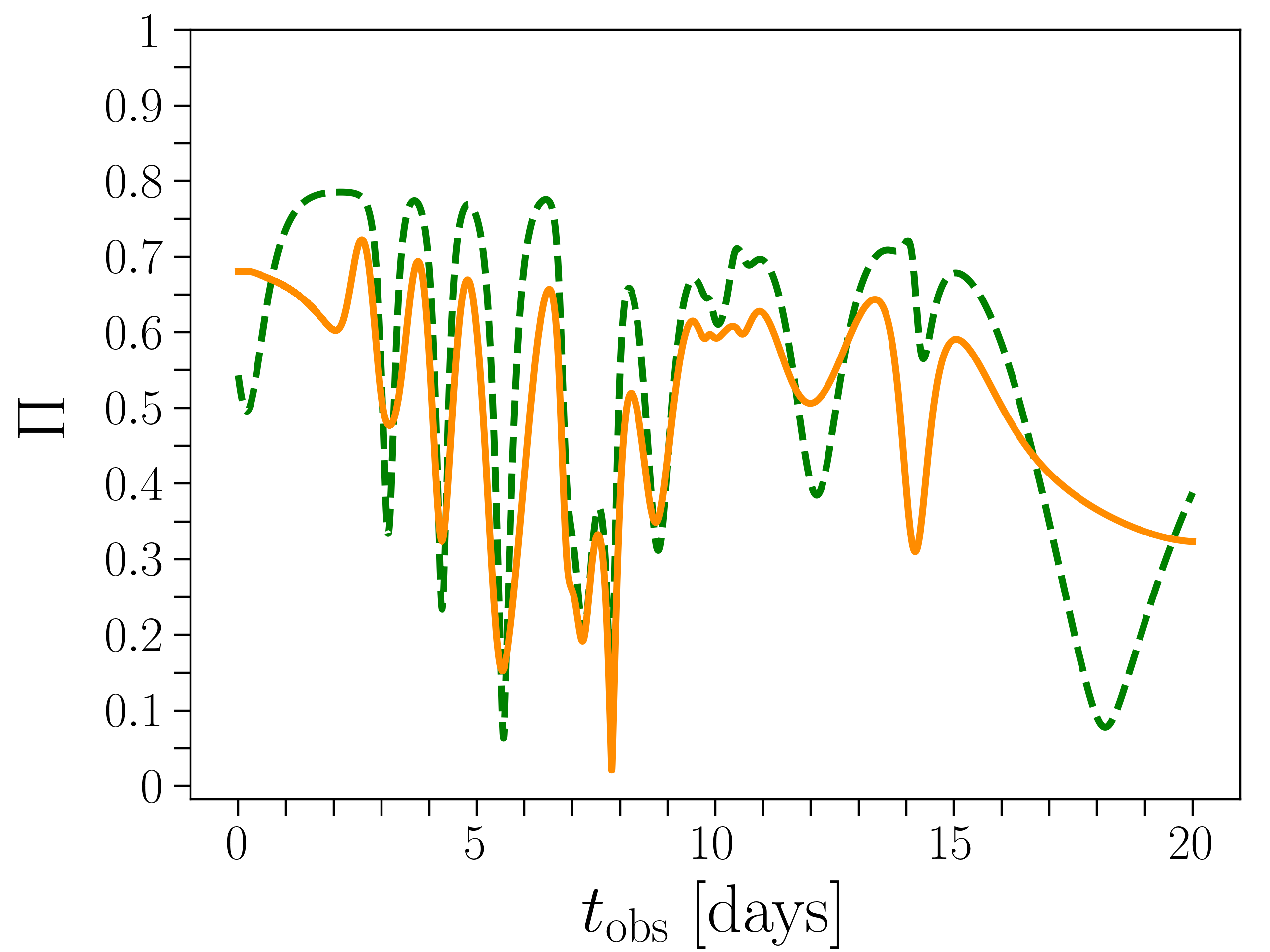}%
    \hspace{+0.25em}%
    \includegraphics[width=0.33\textwidth]{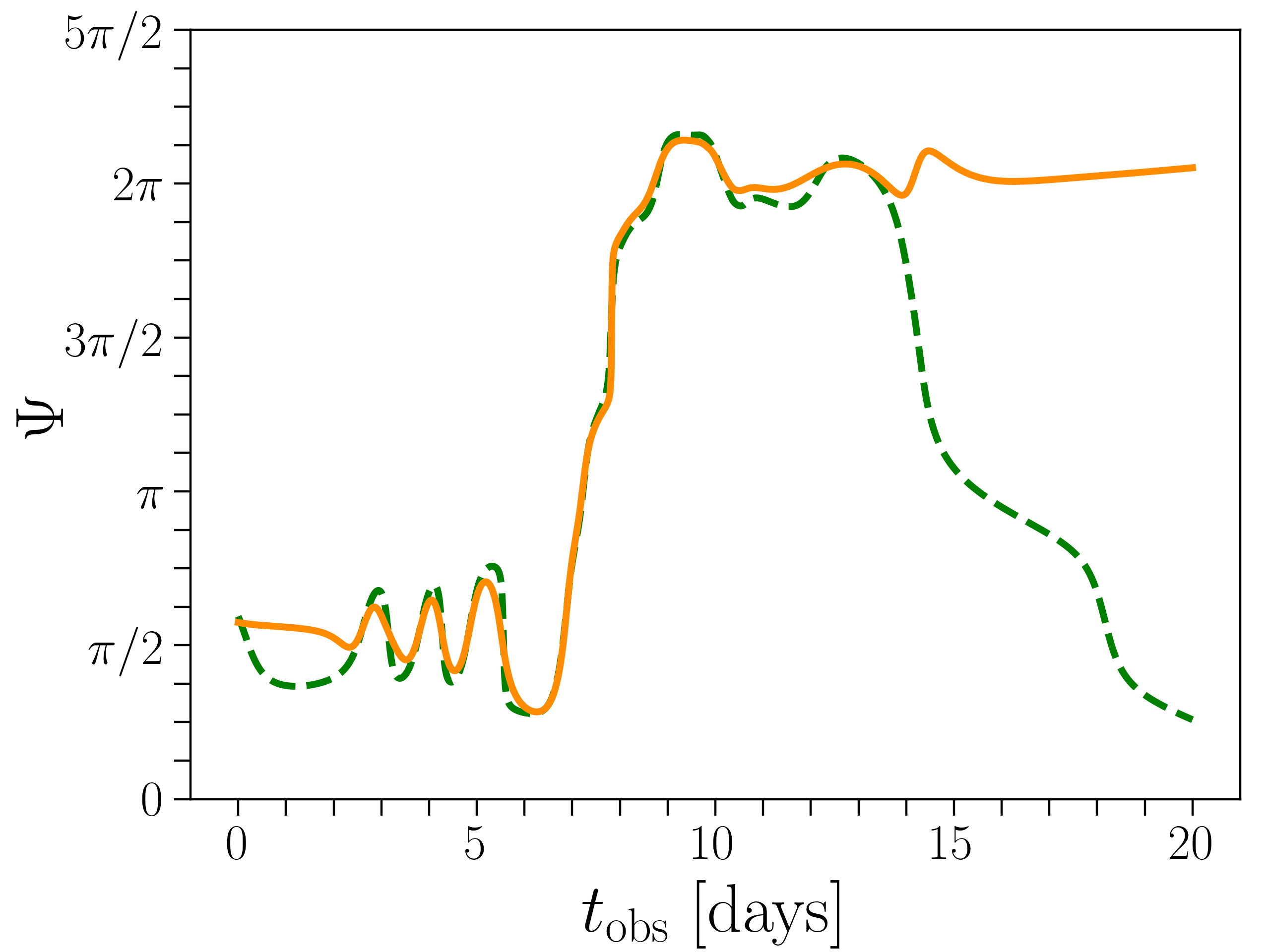}%
    \hspace{+0.25em}%
    \includegraphics[width=0.33\textwidth]{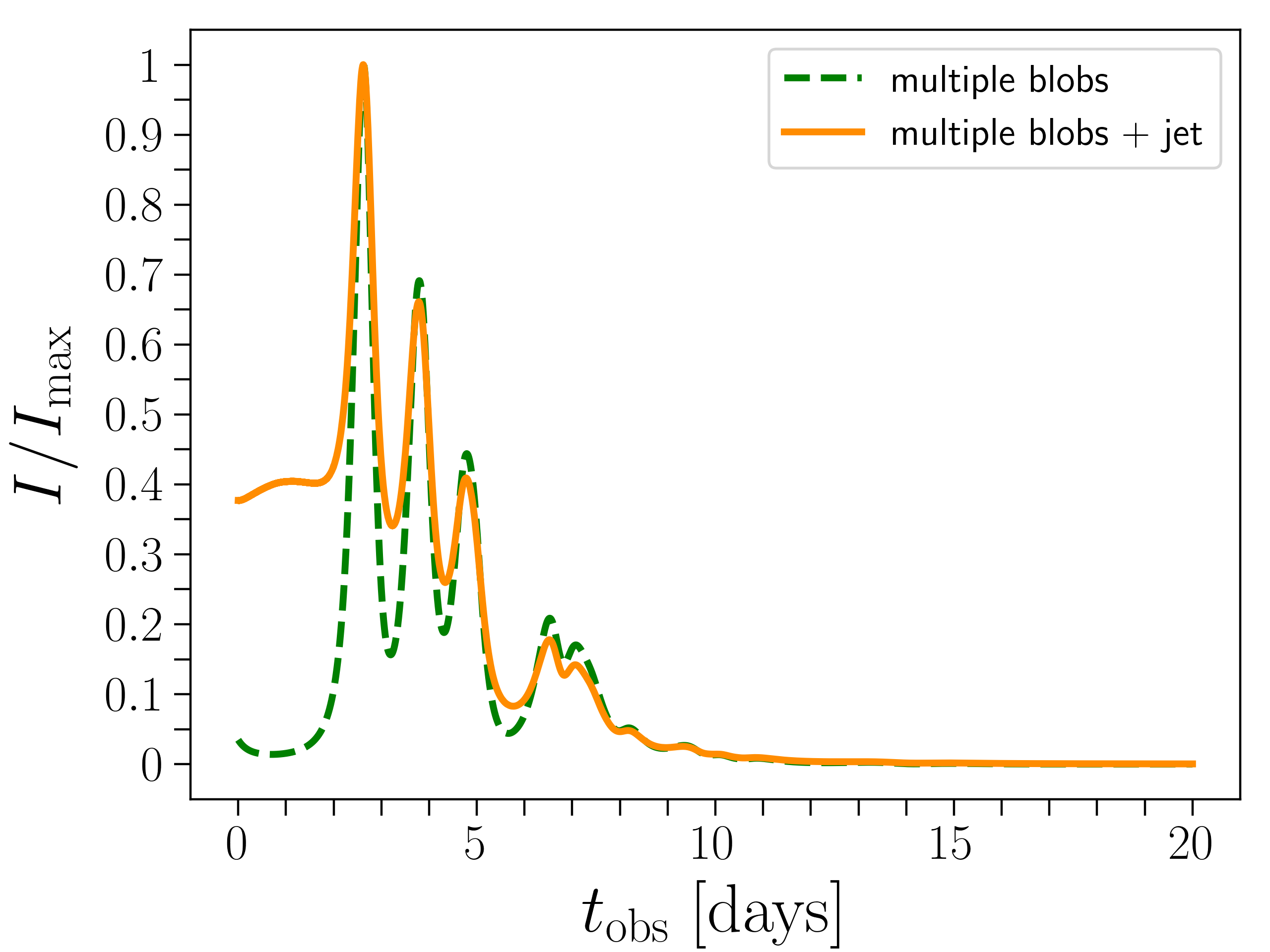}  
    \caption{Same as Fig.~\ref{fig:cilinder}, but the jet shape is nearly parabolic. The first row (``single blob A'') and the second row (``single blob B'') differ only for the initial coordinates of the blob.}
    \label{fig:pl}
\end{figure*}

In Fig.~\ref{fig:pl}, we show the time evolution of $\Pi$ (left panels), $\Psi$ (middle panels), and $I/I_{\rm max}$ (right panels) for a nearly parabolic jet. We consider two scenarios with a single blob (``single blob A'' and ``single blob B'', shown respectively in the top and middle panels). These scenarios differ for the initial coordinates of the blob, which are given in Table \ref{table:param}.

Large EVPA rotations ($\Delta\Psi>180^{\circ}$) may or may not occur depending on the initial coordinates of the blob, which affect its Doppler factor. For the ``single blob A'', when $\eta_{\rm jet}=0$ we observe a rotation of $\Delta \Psi \sim 350^{\circ}$. The fastest rotation rate is associated with a double-peaked intensity structure, with peaks at $t_{\rm obs} \sim 11\;\mathrm{days}$ and $t_{\rm obs} \sim 13\;\mathrm{days}$. For the ``single blob B'', when $\eta_{\rm jet}=0$ we observe a smaller rotation followed by a counter-rotation ($\Delta \Psi \sim 80^{\circ}$), which is associated with a peak in intensity. When $\eta_{\rm jet}=1$, we observe $\Delta \Pi \sim 0.65$ and $\Delta \Psi \sim 118^{\circ}$ for the ``single blob A'', while we observe $\Delta \Pi \sim 0.69$ and $\Delta \Psi \sim 127^{\circ}$ for the ``single blob B''.

In the scenario with multiple blobs (bottom panels), the polarization degree fluctuates substantially, giving the impression of erratic variability, whereas the EVPA rotates more smoothly. We observe large EVPA rotations: $\Delta \Psi \sim 342^{\circ}$ when $\eta_{\rm jet}=0$, and $\Delta \Psi \sim 334^{\circ}$ when $\eta_{\rm jet}=1$. We emphasize that in the case $\eta_{\rm jet}=0$ large EVPA rotations ($\Delta\Psi > 180^{\circ}$) are seen in both directions, clockwise and counterclockwise.

\section{Discussion}
\label{sec:disc}

We investigated the temporal variability of the polarization of the synchrotron radiation from blazar jets. We adopted a geometric and deterministic model. Following \citet{Marscher08}, we considered off-axis emission features (blobs) that propagate along the jet with the local bulk velocity of the fluid. We computed the jet electromagnetic fields and the bulk velocity self-consistently, using the jet model of \cite{Lyubarsky2009}. The jet is axisymmetric, and its electromagnetic fields do not have a turbulent component. We did not assume a specific particle acceleration mechanism. Blobs could form if the acceleration sites are localized, as happens, for example, due to intermittency in a turbulent cascade \citep[][]{ComissoSironi2018, ComissoSironi2019}.

We explored several scenarios corresponding to different jet shapes and different initial spatial configurations of the blobs. We found that the temporal variability of the polarization does not follow a universal trend. Our model shows both smooth, continuous variations and seemingly erratic fluctuations of the polarization degree and the electric vector position angle (EVPA). As noted previously by \citet{Lyutikov2017}, a geometric and deterministic model can produce such features. Interestingly, the same jet shape can produce a different temporal evolution of the polarization degree and EVPA by changing the initial spatial configuration of the blobs.

Our model produces remarkably complex polarization patterns. We observed the following features in the temporal evolution of the polarization degree and EVPA.
\begin{itemize}
\item There is no clear correlation between the polarization degree and the EVPA. During EVPA rotations, the polarization degree can be either constant or
suddenly jump from its minimum to its maximum value.
\item  Simultaneous high-amplitude variations of the polarization degree and the EVPA can coincide with sharp peaks of the observed luminosity.
\item  Bidirectional EVPA rotations (both clockwise and counterclockwise) are possible for the same jet shape. Large EVPA rotations of $180^{\circ}$ or more are occasionally observed.
\item  The polarization degree and the EVPA show significant variability on a broad range of timescales for the same jet shape, from a few hours to several days or even weeks.
\end{itemize}

Our findings are important for interpreting variability signatures in polarimetric observations. The \textit{RoboPol} monitoring program detected several large optical EVPA rotations that are frequently associated with multiwavelength flares \citep[e.g.,][]{Marscher08, marscher10, blinov16, Blinov18}. However, in many cases, the polarization degree and the EVPA patterns are not correlated with flux variability, and their evolution appears to be erratic \citep{marscher14, Kiehlmann2017}.

More recently, the launch of the \textit{IXPE} satellite enabled multiwavelength polarimetric studies of HSP blazars. Interestingly, simultaneous observations of the X-ray and optical EVPA show that their variability is often not correlated. During a five-day monitoring campaign of Mkn 421, \cite{digesu23} reported a $\sim360^{\circ}$ EVPA rotation in the X-ray band, with no corresponding rotation observed at lower frequencies. The polarization degree remained relatively stable throughout the event and showed a strong chromaticity.

\cite{middei23a} reported the first orphan optical EVPA rotation. In PG 1553+113, the optical EVPA changed by approximately $125^{\circ}$, with no simultaneous rotation observed in the X-ray band. The polarization degree showed no substantial variability during the event. \cite{Kouch2024} reported another orphan optical EVPA rotation in PKS 2155-304. The optical EVPA rotated approximately $50^{\circ}$ in one direction and then $40^{\circ}$ in the opposite direction, while the X-ray EVPA remained stable. This behavior demonstrates that clockwise and counterclockwise EVPA rotations can occur within the same source.

During the first \textit{IXPE} pointing of Mkn 421 in May 2022, \cite{Kim2024} detected an X-ray EVPA rotation of approximately $180^{\circ}$, while the X-ray polarization degree was nearly constant. A subsequent EVPA rotation in June 2022 was accompanied by simultaneous spectral variations. Interestingly, at longer wavelengths, the EVPA rotated in the opposite direction with respect to the X-ray EVPA, and the rotation occurred over a longer timescale.

During an optical–X-ray flaring episode of the LSP blazar S4 0954+65, \cite{Kouch2025} observed two orphan optical EVPA rotations: one of approximately $85^{\circ}$ that lasted one day and another of $\sim 87^{\circ}$ that lasted two days. In contrast, during the same flaring episode the X-ray polarization degree varied by a factor of two, whereas the optical polarization degree was stable.

According to the standard interpretation of the \textit{IXPE} collaboration \citep{digesu23}, temporal variability of polarization indicates that non-thermal particles are accelerated by shocks that move on a helical trajectory. This interpretation explains orphan X-ray EVPA rotations because the accelerated particles are energy stratified. The optical EVPA does not rotate because the emitting electrons are distributed more uniformly as a result of their longer cooling time. However, this model does not explain orphan optical EVPA rotations.

We remark that in our model we do not prescribe the energy distribution of the electrons in the blobs or in the jet. Therefore, the time evolution that we derive is generic (i.e.,~it does not specifically apply to a particular band and can thus be used for both optical and X-ray observations). It is plausible that the spectral properties of the blobs and the background are different. This would add more complexity to the scenario. Indeed, the observation of both optical and X-ray EVPA rotations indicates that the ratio of the background emission (produced from a nearly axisymmetric jet) and the emission from the blobs can be strongly chromatic. Large EVPA rotations can be produced when the blobs dominate the observed emission, while more erratic variability is expected when the emission from the jet is larger. It is not clear whether the particle acceleration mechanism could be constrained on the basis of these relatively straightforward considerations. As discussed previously by \citet{Bolis+2024, Bolis+2024b}, constraining the acceleration mechanism from multifrequency polarimetric observations remains extremely challenging.

\begin{acknowledgements}
We thank the anonymous referee for their constructive comments that improved the paper. We acknowledge financial support from a Rita Levi Montalcini fellowship (PI E.~Sobacchi) and an INAF Theory Grant 2024 (PI F.~Tavecchio). This work has been funded by the European Union-Next Generation EU, PRIN 2022 RFF M4C21.1 (2022C9TNNX).
\end{acknowledgements}

\bibliographystyle{aa}
\bibliography{tavecchio}

\appendix

\section{Stokes parameters of a single blob}
\label{sec:StokesT}

In the following, we describe our method to calculate the Stokes parameters of synchrotron radiation emitted by a single, compact emission feature (``blob'') that moves with the local drift velocity of the MHD flow. The Stokes parameters can be presented as \citep{DelZanna06}
\begin{align}
\label{eq:stokes1}
I & = \frac{p + 7/3}{p+1}  \int {\rm d}V  \;  j \left(t, \mathbf{r} \right) \\
\label{eq:stokes2}
Q & =  \int {\rm d}V \;  j \left(t, \mathbf{r} \right)  \,\cos 2 \chi \\
\label{eq:stokes3}
U & = \int {\rm d}V  \;  j \left(t, \mathbf{r} \right) \, \sin 2 \chi \, ,
\end{align}
where the integration volume is ${\rm d}V = R\; {\rm d}R \; {\rm d}\phi\; {\rm d}z$. When the energy distribution of the electrons can be described as a power law, the emission coefficient, $j$, is given by \citep{Lyutikov2003, DelZanna06, Bolis+2024}
\begin{equation}
j = \kappa_p \, K_{\rm e}\left(t, \mathbf{r} \right) \, \mathcal{D}^{(3 + p)/2} \, \left| \mathbf{B}' \times \hat{\mathbf{n}}' \right|^{(p + 1)/2}\;,
\end{equation}
where $K_{\rm e}$ is the proper density of the electrons in the blob, $\mathcal{D}=[\Gamma (1-\mathbf{v}\cdot\hat{\mathbf{n}})]^{-1}$ is the Doppler factor, $|\mathbf{B}' \times \hat{\mathbf{n}}'|$ is the strength of the magnetic field component perpendicular to the line of sight measured in the frame of the fluid. The polarization angle, $\chi$, denotes the angle between the polarization vector of synchrotron radiation produced by a volume element and the projection of the jet axis on the plane of the sky. The constant $\kappa_p$, which does not affect the polarization degree and the EVPA, is defined in Eqs.~(12)-(13) of \citet{Bolis+2024}.

Following \citet{DelZanna06}, one can express the quantities that appear in Eqs.~\eqref{eq:stokes1}-\eqref{eq:stokes3} as functions of the electromagnetic fields measured in the observer's frame \citep[for a detailed derivation, see][]{Bolis+2024, Bolis+2024b}. The unit vector directed toward the observer is $\hat{\mathbf{n}}= \sin{\theta_{\rm obs}}\cos{\phi}\; \hat{\mathbf{R}} - \sin{\theta_{\rm obs}} \sin{\phi}\; \hat{\bm{\phi}}+ \cos{\theta_{\rm obs}}\hat{\mathbf{z}}$, where the viewing angle, $\theta_{\rm obs}$, is measured with respect to the direction of the jet axis, $\hat{\mathbf{z}}$. The polarization vector of the radiation from a volume element is given by \citep{Lyutikov2003, DelZanna06}
\begin{equation}
\label{eqn:polvec}
\hat{\mathbf{e}} = \frac{\hat{\mathbf{n}} \times \mathbf{q}}{\sqrt{  q^{2} - \left( \mathbf{q} \cdot \hat{\mathbf{n}} \right)^{2} }} \;,  \qquad \mathbf{q} = \mathbf{B} - \hat{\mathbf{n}} \times \mathbf{E} \;.
\end{equation}
The polarization angle, $\chi$, is given by $\cos\chi=\hat{\mathbf{e}} \cdot \hat{\mathbf{l}}$ and $\sin\chi=\hat{\mathbf{e}} \cdot (\hat{\mathbf{l}} \times \hat{\mathbf{n}})$, where $\hat{\mathbf{l}}=[(  \hat{\mathbf{z}} \cdot \hat{\mathbf{n}})\hat{\mathbf{n}}-\hat{\mathbf{z}}]/\sqrt{1-(\hat{\mathbf{z}} \cdot \hat{\mathbf{n}})^2}$ is the projection of the jet axis on the plane of the sky. The final expressions for $\mathcal{D}$, $\left| \mathbf{B}' \times \hat{\mathbf{n}}' \right|$, $\chi$ as a function of $\mathbf{E}$, $\mathbf{B}$ are given by Eqs.~(21)-(23) of \citet{Bolis+2024}.

To study the time variability of the Stokes parameters, it is important to express the time when photons are emitted, $t$, as a function of the time when they are observed, $t_{\rm obs}$. The coordinates of the emitter are
\begin{equation}
\mathbf{r}=\left(R\cos\phi,\; R\sin\phi,\; z\right)
\end{equation}
and the coordinates of the observer are
\begin{equation}
\mathbf{r_{\rm obs}} = \left(d\sin \theta_{\rm obs},\;  0,\;  d\cos \theta_{\rm obs} \right) \;.
\end{equation}
Taking into account that $d \gg R_{\rm blob}, z_{\rm blob}$ (i.e.,~the observer is effectively at infinity), the distance that a photon must travel to reach the observer can be approximated as
\begin{equation}
\left| \mathbf{r_{\rm obs}}  - \mathbf{r} \right| = d -  R \cos \phi \sin \theta_{\rm obs} - z \cos \theta_{\rm obs} \;.
\end{equation}
The corresponding time delay can be presented as
\begin{equation}
\label{eq:time}
t = t_{\rm obs} +  R \cos \phi \sin \theta_{\rm obs} + z \cos \theta_{\rm obs} \;.
\end{equation}
We neglected the constant offset, $d$, which has no effect on relative arrival times.

The blob moves with the local drift velocity of the flow. Its coordinates, $\mathbf{r}_{\rm blob}(t)$, are governed by the equation of motion ${\rm d}\mathbf{r}_{\rm blob}/{\rm d}t=\mathbf{E}\times\mathbf{B}/B^2$. To calculate the Stokes parameters given by Eqs.~\eqref{eq:stokes1}-\eqref{eq:stokes3}, one should specify the proper number density of the electrons, $K_{\rm e}(t, \mathbf{r})$. We assume that
\begin{equation}
\label{eq:emiss}
K_{\rm e} = K_0 \; F\!
\left( \frac{R - R_{\rm blob}}{ \Delta R} \right) F\!\left( \frac{\phi - \phi_{\rm blob}}{ \Delta R / R_{\rm blob}} \right) F\!\left( \frac{z - z_{\rm blob}}{ \Delta R / \Gamma} \right)\,,
\end{equation}
where $K_0$ is a constant, $\Gamma$ is the Lorentz factor of the blob, and $\Delta R$ characterizes its spatial extent. Since the blob is compact, we have $\Delta R\ll R_{\rm blob}$. The function $F$ is defined as
\begin{equation}
F\!(\xi )= 
\begin{cases}
    1 & \text{if } \, 0<\xi<1 \\
0 & \text{otherwise}\,.
\end{cases}
\end{equation}
The factor $\Delta R/\Gamma$ in the $z$-term of Eq.~\eqref{eq:emiss} accounts for length contraction in the direction of motion (in the model presented in Sect.~\ref{sec:jetmodel}, one has $v_z \gg v_R, v_\phi$). Now, it is convenient to make the following change of variables
\begin{align}
\widetilde{R} & =  \frac{R - R_{\rm blob}}{\Delta R}  \\
\widetilde{\phi} & = \frac{\phi - \phi_{\rm blob}}{\Delta R / R_{\rm blob}} \\
\widetilde{z} & =  \frac{z - z_{\rm blob}}{ \Delta R / \Gamma}  \;.
\end{align}
The determinant of the corresponding Jacobian matrix, $\mathfrak{J}$, can be approximated as
\begin{equation}
{\rm det}\left(\mathfrak{J}\right) = \frac{\left(\Delta R \right)^3\mathcal{D} }{R_{\rm blob}}\;.
\end{equation}
The Stokes parameters are given by
\begin{align}
\label{eq:Iblob}
I &= \frac{p + 7/3}{p+1} \, \left( \Delta R \right)^3 \, \mathcal{D}
    j_0  \\
\label{eq:Qblob}
Q & = \left( \Delta R \right)^3 \, \mathcal{D}\,
    j_0  \, \cos 2 \chi \\
\label{eq:Ublob}
U & = \left( \Delta R \right)^3 \, \mathcal{D}\,
    j_0  \, \sin 2 \chi \,,
\end{align}
where $ j_0 = \kappa_p \,K_0\mathcal{D}^{\left( 3 + p\right)/2} \,  \left| \mathbf{B}' \times \hat{\mathbf{n}}' \right|^{\left( p+1 \right)/2}$. To determine the Stokes parameters as a function of $t_{\rm obs}$, one should evaluate the physical quantities in Eqs.~\eqref{eq:Iblob}-\eqref{eq:Ublob} at the position $\mathbf{r}_{\rm blob}(t)$. Then, one should express $t$ as a function of $t_{\rm obs}$ using Eq.~\eqref{eq:time}.

\section{Summary of results}

In Table~\ref{table:results} we summarize our results.

\renewcommand{\arraystretch}{1.5}
\begin{table*}
\caption{\label{table:results} Polarization degree and EVPA for different jet shapes and emission scenarios.}
\centering
\begin{tabular}{l|l|ccc|ccc}
\toprule
Jet shape 
& Emission scenario
& $\Pi_{\rm min}$ 
& $\Pi_{\rm max}$ 
& $\Delta \Pi$ 
& $\Psi_{\rm min}$ 
& $\Psi_{\rm max}$ 
& $\Delta \Psi$ \\
\midrule

\multirow{4}{*}{Cylindrical} 
  & single blob           & 0.789 & 0.789 & 0     & 1.050 & 2.092 & 1.042 \\
  & single blob + jet     & 0.610 & 0.782 & 0.172 & 1.190 & 1.952 & 0.762 \\
  & multiple blobs        & 0.412 & 0.788 & 0.376 &  1.077 & 2.046 & 0.969 \\
  & multiple blobs + jet  & 0.578 & 0.738 & 0.160 & 1.348 & 1.778 & 0.430 \\
\midrule

\multirow{4}{*}{\shortstack{``Sausage-like''\\ $C_2=0.8$}} 
  & single blob           & 0.789 & 0.789 & 0     & 0.909 & 2.262 & 1.353 \\
  & single blob + jet     & 0.520 & 0.782 & 0.262 & 1.081 & 2.163 & 1.082 \\
  & multiple blobs        & 0.238 & 0.788 & 0.550 & 0.900 & 2.254 & 1.354 \\
  & multiple blobs + jet  & 0.493 & 0.741 & 0.248 & 1.317 &  1.966 & 0.649 \\

\multirow{4}{*}{\shortstack{``Sausage-like''\\ $C_2=2$}} 
  & single blob           & 0.789 & 0.789 & 0     & -9.061 & 5.321 & 14.382 \\
  & single blob + jet     & 0.170 & 0.766 & 0.596 & 0.876  &  2.345 & 1.469 \\
  & multiple blobs        & 0.023 & 0.779 & 0.756 & -4.672 & 2.192 & 6.864 \\
  & multiple blobs + jet  & 0.034 & 0.654 & 0.620  & 0.832  & 2.283 & 1.451 \\
\midrule

\multirow{6}{*}{Nearly parabolic} 
  & single blob A         & 0.789 & 0.789 & 0     & -1.083 &  5.004 & 6.087 \\
  & single blob A + jet   & 0.121 & 0.777 & 0.656 & -0.259  &1.805 & 2.064 \\
  & single blob B         & 0.789 & 0.789 & 0     & 0.998  & 2.443 & 1.445 \\
  & single blob B + jet   & 0.075 &  0.769 & 0.694 & 0.135  & 2.360 & 2.225 \\
  & multiple blobs        & 0.063 & 0.785 & 0.722 & 0.810  &  6.786 & 5.976\\
  & multiple blobs + jet  & 0.021 & 0.722 & 0.701 & 0.890  & 6.727 & 5.837 \\
\bottomrule
\end{tabular}
\tablefoot{
We report the minimum and maximum values of the polarization degree, $\Pi$, and EVPA, $\Psi$ (measured in radians), for each jet shape and emission scenario, over the entire observation period, $t_{\rm obs} = 20{\rm\; days}$. We define $\Delta \Pi=\Pi_{\rm max}-\Pi_{\rm min}$ and $\Delta \Psi = \Psi_{\rm max}-\Psi_{\rm min}$. The time evolution of the cylindrical jet is shown in Fig.~\ref{fig:cilinder}, the time evolution of the ``sausage-like'' jet is shown in Figs.~\ref{fig:sausageA}--\ref{fig:sausageB}, and the time evolution of the nearly parabolic jet is shown in Fig.~\ref{fig:pl}. When the jet is nearly parabolic, ``single blob A'' and ``single blob B'' correspond to different initial coordinates of the blob, as reported in Table \ref{table:param}.
}
\end{table*}

\end{document}